\documentclass{article}

\usepackage{arxiv}

\usepackage[utf8]{inputenc} % allow utf-8 input
\usepackage[T1]{fontenc}    % use 8-bit T1 fonts
\usepackage{hyperref}       % hyperlinks
\usepackage{url}            % simple URL typesetting
\usepackage{booktabs}       % professional-quality tables
\usepackage{amsfonts}       % blackboard math symbols
\usepackage{nicefrac}       % compact symbols for 1/2, etc.
\usepackage{microtype}      % microtypography
\usepackage{lipsum}
\usepackage{array}
\usepackage{graphicx}
\usepackage{caption}
\usepackage{subcaption}
\usepackage{epstopdf, epsfig}
\usepackage{amsmath}
\usepackage{comment}
\usepackage{textcomp}

\usepackage{enumitem}
\usepackage{upgreek,bm}
\usepackage{booktabs}
\usepackage{siunitx}
\usepackage{gensymb}
\usepackage[space]{grffile}
\usepackage{float}

\title{On the physical interpretation of proper orthogonal decomposition and dynamic mode decomposition for liquid injection}

\author{
  Scott B.~Leask\thanks{Preprint, work in progress.} \\
  Department of Mechanical and Aerospace Engineering\\
  University of California, Irvine\\
  Irvine, CA, 92697, USA \\
  \texttt{sbl@ucicl.uci.edu} \\
   \And
  Vincent G.~McDonell \\
  Department of Mechanical and Aerospace Engineering\\
  University of California, Irvine\\
  Irvine, CA, 92697, USA \\
  \texttt{mcdonell@ucicl.uci.edu} \\\\  
}

\begin{document}
\maketitle

\begin{abstract}
The modal decomposition techniques of proper orthogonal decomposition (POD) and dynamic mode decomposition (DMD) have become a common method for analysing the spatio-temporal coherence of dynamical systems. In particular, these techniques are of interest for liquid injection systems due to the inherent complexity of multiphase interactions and extracting the underlying flow processes is desired. Although numerous works investigating flow processes have implemented POD and DMD, the results are often highly interpretive with limited link between the decomposition theory and the interpreted physical meaning of the extracted modes. Here, we provide insight into the interpretation of POD and DMD modes in a hierarchical structure. The interpretation of modes for simple canonical systems is validated through knowledge of the underlying processes which dominate the systems. We show that modes which capture true underlying phenomena produce subsequent modes at higher harmonics, up until the Nyquist limit, whose modal structure scales decrease proportionally with increasing modal frequency. These higher harmonics primarily encode motion information and may or may not capture additional structural information, which is dependent on the system. We demonstrate these findings first on canonical liquid injection systems to enhance the interpretation and understanding of results extracted from practical jet in crossflow systems.
\end{abstract}

\section{Introduction}\label{Introduction}

Combustion and fluid flow problems are typically very difficult to model accurately and generally. Difficulties arise due to certain characteristics which are inherently found in these systems; the systems exhibit high-dimensionality, complex coupling of system processes, and often suffer from noise or incomplete data \cite{lefebvre2017atomization}. In a practical experiment, system processes abound. A variety of procedures can be taken to isolate underlying system processes or to limit the effect of undesirable processes, however the recorded data will still likely be an entangled combination of numerous process modes or behaviours. Even the simplest of these systems lack analytical theory based off first principles as multiphase interactions and turbulence greatly inhibit detailed physical interactions \cite{bradshaw1994turbulence}. Where analytic theory is hindered, researchers have the opportunity to learn directly from the system probed itself. 

Large quantities of data are becoming ubiquitous in all science and engineering fields. As a result, learning from these big data collections through data-driven models has become a serious undertaking to facilitate system analysis and to understand low-dimensional representations of high-dimensional systems. With the growth of big data, machine learning has flourished and is quickly becoming a popular tool in solving heretofore difficult or intractable problems throughout science and engineering. Although impressive progress has been made in addressing problems through machine learning techniques, much of the current progress is merely a means to an end; solving a problem is prioritised over understanding the problem itself. A better understanding and interpretation of how these models learn has the potential to aid understanding of the systems analysed, to improve how the models solve problems, and to aid understanding in how the models generalise, the latter of which is of great interest \cite{zhang2016understanding}.

Modal decomposition techniques are currently the most prominent data-driven method in the fluid mechanics community which, too, have seen more widespread interest. Many of these techniques are founded in abstract theory whose application purports to extract physical understanding yet the extracted results are not fully understood and validated. In a fluid system, for example, although a large number of modes may be affecting the flow dynamics, it is often the case that the flow is dominated by only a small subset of these modes and that an accurate reconstruction of the entire system may be computed with knowledge of only a few modes. Dimensionality reduction can then be performed, whereby insignificant modes are removed based on, for example, an energy content metric, which can reduce redundancy in datasets and illuminate dominant flow processes \cite{van2009dimensionality, herrmann2012fighting}.

Proper orthogonal decomposition (POD) and dynamic mode decomposition (DMD) are two modal decomposition approaches to perform this dimensionality reduction. The main idea of these techniques is to extract coherent spatio-temporal modes which dominate the measured system. These modes can be used for predictive modeling, data compression, non-linear control, or a means to understand and extract the governing physics of the system.

The application of POD and DMD can be found extensively in works on fluid flow phenomena. Early applications of POD come from Berkooz \textit{et al.} \cite{berkooz1993proper} and Kirby \textit{et al.} \cite{kirby1990proper} for the analysis of turbulent flows and supersonic shear layers, respectively. These two works represent different approaches to POD, with the latter using the, now, more widely used snapshot POD approach introduced by Sirovich \cite{sirovich1987turbulence}. The seminal work on DMD was by Schmid  \cite{schmid2010dynamic} who highlights differences between POD and DMD and provides example applications and whose follow up work gives insight into extracted representative dynamic modes for fluid flow cases \cite{schmid2011applications}.

More specifically, POD and DMD are receiving increasing attention in the area of liquid injection and atomisation. Theory governing liquid injection and subsequent near-field and far-field behaviour is limited to very simple, non-practical cases which do not generalise well. As atomisation problems are inherently complex, yet demonstrate wave-like oscillatory transport and breakup mechanisms \cite{lefebvre2017atomization}, POD and DMD appear well suited to develop an understanding of underlying processes which control the dynamical system. These techniques have been applied to numerous systems of interest. For example, recently for POD, Charalampous and Hardalupas \cite{charalampous2014application} investigated the morphology of immiscible liquids, Gadiraju \textit{et al.} \cite{gadiraju2017application} used POD to aid flame dynamic understanding from an industrial fuel nozzle, De Giorgi \textit{et al.} \cite{de2018characterization} identified cavitation processes in a sharp-edged orifice, and Taghizadeh and Jarrahbashi \cite{taghizadeh2018proper} captured vorticity structures governing the flow of cryogenic liquids injected at transcritical conditions.

For DMD applications, Rowley \textit{et al.} \cite{rowley2009spectral} applied DMD to 3D data of a simulated jet in crossflow and Leask \textit{et al.} \cite{leask2019emulsion} applied DMD to experimental data of an emulsion jet in crossflow. Further works include the analysis of cyclic variations of a pulsing spray \cite{chen2013analyzing}, internal flow of a pressure swirl atomiser \cite{vashahi2018large}, and external flow of a pressure swirl atomiser \cite{leask2019preliminary}.

A number of works have investigated the differences between the POD and DMD techniques. Tirunagari \textit{et al.} \cite{tirunagari2012analysis} applied POD and DMD to data from a large eddy simulation of injected gaseous jets at subsonic conditions and were able to compare the techniques' efficacy at identifying Kelvin-Helmholtz instability. Magionesi \textit{et al.} \cite{magionesi2018modal} performed a comparative analysis of POD and DMD in capturing the dynamic evolution of flow past a marine propeller, Torregrosa \textit{et al.} \cite{torregrosa2018modal} noted differences between these techniques in extracting acoustic resonance behaviours from a compression-ignited combustion chamber, and Muld \textit{et al.} \cite{muld2012flow} extracted coherent flow structures in the wake of a high-speed train. Hua \textit{et al.} \cite{hua2016dynamic} investigated the application of DMD on shear coaxial jets which looked into the robustness of extracted modes, following on work by Roy \textit{et al.} \cite{roy2015deconvolution}. These works note the difficulty in interpreting POD and DMD modes but with their respective advantages and, in the case of Hua \textit{et al.} \cite{hua2016dynamic}, acknowledge the lack of verification in successfully differentiating true underlying modes from system noise.

Certainly, a large number of works have applied POD, DMD, or both in parallel, especially to non-linear systems. Interestingly, in certain cases, the agreement of DMD with POD is used as validation that the modal decompositions are extracting meaningful results, irrespective of their close relationship in theory. Although POD and DMD are becoming more prevalent in the literature, few works focus on the physical interpretation of extracted modes, seek to validate this interpretation, or identify decomposition dependence to user parameters, such as the number of frames to analyse. As a result, there is a risk that interpretations of results, or the results themselves, do not represent true underlying system processes.

Some works have made attempts to understand physically what the extracted modes mean. Physical interpretation of POD modes has been discussed by Kerschen and Golinval \cite{kerschen2002physical} for structural dynamics problems where a focus on linear systems revealed that POD will at least converge to the system eigenmodes. For the non-linear case tested, however, no intuition was given on what the best linear representation means physically. Similarly, Feeny and Liang \cite{feeny2003interpreting} interpreted POD modes of a linear vibration system by correlating them with the true vibrational modes under random excitation. Indeed, good agreement in the POD modes and the normal modes of the system have been found for linear systems, but greater difficulty arises through the linearisation of non-linear systems.

Chen \textit{et al.} \cite{chen2012use, chen2013practical} noted the difficulty in interpreting POD modes for non-linear systems, particularly when comparing likeness between two datasets as POD energy distributions or spectra are not a good correlation metric. Further, they demonstrate the highly-coupled nature of processes in systems as all flow structures in every data matrix snapshot contributes, to some extent, to every POD mode. Arienti and Soteriou \cite{arienti2009time} did a thorough look of POD for liquid injection cases. While a large number of modes and their interpretations are given, this is not the case for all modes where some are left out of the discussion. As is commonly found in practical applications, certain modes are particularly easy to discuss, while others are not. In addition, distinguishing between a valid mode or noise can be a significant challenge \cite{dawson2016characterizing}, particularly when the data are resolution-limited. The majority of interpretations given in works is done without too much, or any, validation. The work we present here can be considered an extension of the previous work by Arienti and Soteriou \cite{arienti2009time} but with an emphasis on DMD and interpretation validation.

Oftentimes, image-based recordings of a dynamical system are used for modal decomposition as they are a convenient and non-intrusive approach to capture a large spatial extent of the system at a, potentially, very high time-resolution. For liquid injection systems, workers are greatly limited in their diagnostic tools due to complex multiphase interactions and instabilities which plague both the near- and far-field areas. This can often exclude commonly used techniques, such as (time-resolved) particle image velocimetry (PIV). Instead, learning physical structures through high-speed video offers a convenient way of extracting very high spatial- and temporal-resolution data from the probed system. Further, the field of computer vision has already seen large improvements in capabilities from learning from pixel-value data alone \cite{parker2010algorithms}. Of course, high-speed video captures a two-dimensional representation of an, often, three-dimensional dynamical system but this does not prevent meaningful results from being extracted. Three-dimensional imaging techniques may provide several advantages, for example through digital holography \cite{schnars2015digital}, but we do not consider their use here due to their limitations on spatial-resolution or setup.

Of the large number of papers which have implemented both POD and DMD there is still not a sound understanding of what these modes represent and how to interpret them in a validated way for non-linear systems. Interpretative claims, in most cases, have not been validated which hinders the progression of understanding flow processes. Therefore, a need exists for work which provides a systematic validation procedure for understanding and interpreting the results outputted by both POD and DMD for liquid injection problems. In addition, there is still a need to compare these techniques to highlight their respective strengths. This paper seeks to address these needs by first investigating simple canonical systems with known characteristics. This allows for sound interpretation and understanding of the extracted modes and corresponding modal information which then guides the interpretation of more complex systems. This then develops until a practical spray may be analysed with a deeper understanding of what the decomposition methods extract.

A number of liquid injection cases are investigated in this work, but it is hoped that the conclusions gathered can be generalised to other systems of interest. The hierarchical validation process is designed to highlight particular traits of POD and DMD that are inherent to the algorithmic behaviour, rather than being system-dependent. Although this work focuses on liquid injection, there are likely correlations that can be made to fluid mechanics problems in general.

It is important to note that this work is not intended to be a deep dive on the theory of either POD or DMD, but rather an attempt to provide insight into the intuition of how these techniques work and what their results represent.

In \S\ref{Intuition}, we discuss the intuition behind the theory of POD and DMD to clarify notable attributes of the techniques. We cover the details of the systems analysed and the validation methodology in \S\ref{methodology} before investigating canonical flow systems with known behaviours in \S\ref{known} and practical jet in crossflow systems with unknown behaviours in \S\ref{unknown}.
 
\section{Decomposition intuition}\label{Intuition}

In this work, only proper orthogonal decomposition (POD) and dynamic mode decomposition (DMD) are considered, where for the latter we use the standard formulation provided by, for example, Tu \textit{et al.} \cite{tu2014dynamic}. Although many variants of these techniques are available in the literature, the standard POD and DMD algorithms are by far the most prevalent and continue to be used in current works. The recent overview of relevant data-driven techniques by Brunton and Kutz \cite{brunton2019data} covers intuition of both POD and DMD which we expand on here. Further, a deeper understanding of these standard algorithms can have an impact on all current and future variants. A brief overview of some common variants is given in \S\ref{PODDMDvariants}.

\subsection{Understanding POD}\label{PODintuition}

Proper orthogonal decomposition has a long history, going back to 1901 with Pearson \cite{peason1901lines} who established principal component analysis (PCA). Since then, PCA developed into the Karhunen-Lo\`eve transform, the Hotelling transform, empirical orthogonal functions, and then Lumley \cite{lumley1967structure} provided the first introduction to POD. All of these techniques are approximately identical, up to nomenclature, mean-subtraction, and variance normalisation variations. For POD, while mean-subtraction does not have an effect on the basic calculations, it can affect the interpretation of the results \cite{chatterjee2000introduction}. We acknowledge that it is typical to subtract the mean for performing POD, however, we do not follow this convention as this allows the mean flow of the system to be captured which provides a valuable reference when comparing extracted modes. In effect, these techniques can all be calculated from the singular value decomposition (SVD) which can be applied to any arbitrary data matrix, $\bm{\mathsf{X}}\in\mathbb{C}^{m\times n}$, to give
\begin{equation}
\bm{\mathsf{X}}=\bm{\mathsf{U}}\bm{\mathsf{\Sigma}}\bm{\mathsf{V}}^\ast\label{PODsvd}
\end{equation}
where $\bm{\mathsf{U}}\in\mathbb{C}^{m\times n}$ has orthonormal columns, $\bm{\mathsf{\Sigma}}\in\mathbb{C}^{n\times n}$ is a diagonal matrix, and $\bm{\mathsf{V}}\in\mathbb{C}^{n\times n}$ has orthonormal columns, and the asterisk denotes the complex conjugate transpose. The columns of the matrix $\bm{\mathsf{U}}$ are typically denoted the POD modes, however, this is not necessarily the case. This formulation is elegantly simple, but some intuition is lost if the POD modes are found blindly from the SVD.

\begin{figure}
    \centering
    \includegraphics[width=0.4\textwidth]{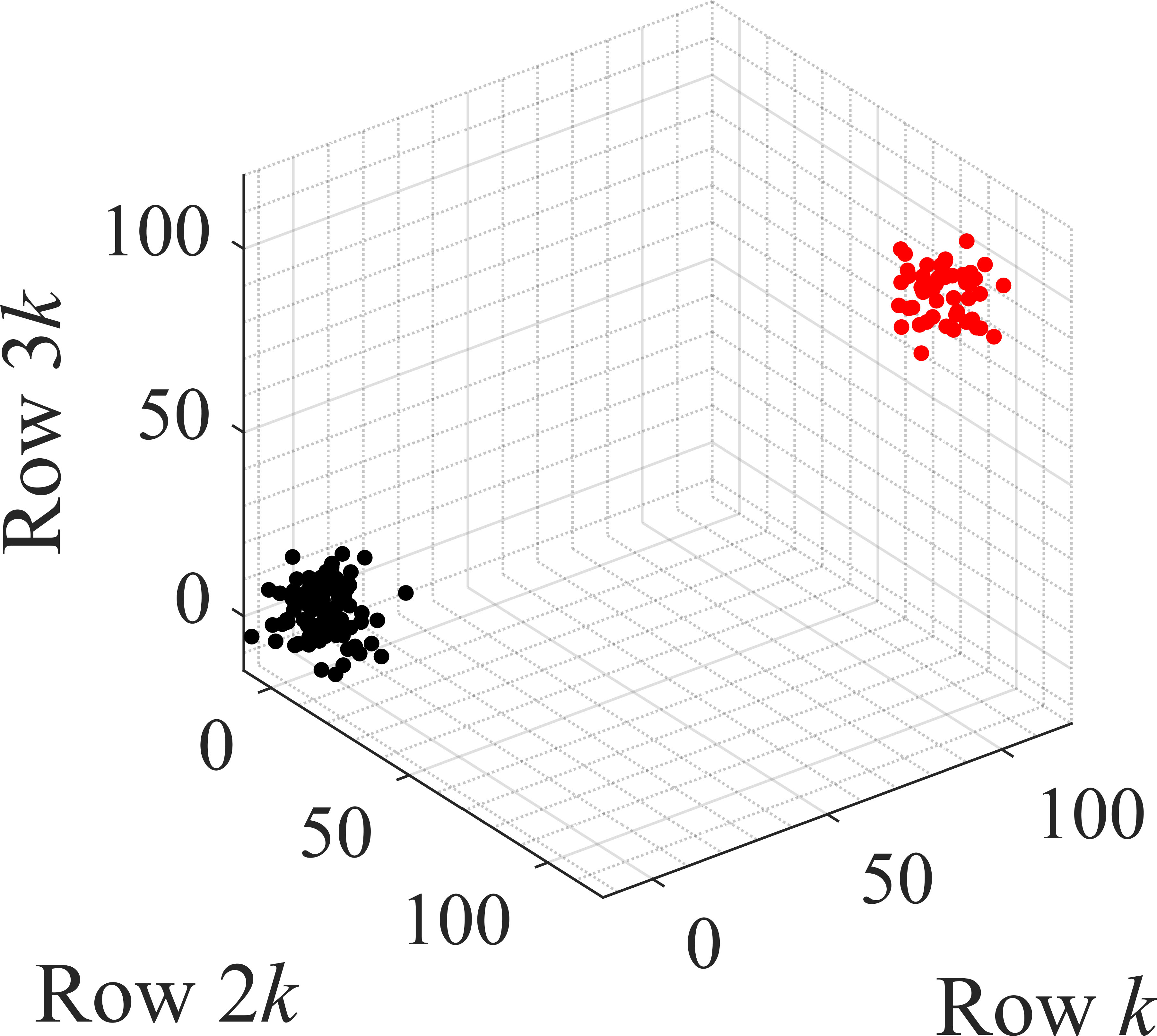}
    \caption{Plot of row similarities. Red points show how every $k^{\text{th}}$ row is highly correlated, but generally uncorrelated with all other rows, given by the black points.}
    \label{fig:covar explain}
\end{figure}

Understanding the data matrix and how the SVD is calculated aids the intuition of how and why POD works. First, the matrix $\bm{\mathsf{U}}$ is calculated as the eigenvectors of the matrix $\bm{\mathsf{X}}\bm{\mathsf{X^\ast}}\in\mathbb{C}^{m\times m}$. The matrix multiplication $\bm{\mathsf{X}}\bm{\mathsf{X^\ast}}$ yields what is essentially a correlation matrix: each row of $\bm{\mathsf{X}}$ is dotted with every other row, including itself, which provides a similarity measure. Therefore, column $j$ and row $j$ of the correlation matrix both describe the similarity of the $j^{\text{th}}$ row of the data matrix with every other row. The correlation matrix can now be thought of as $m$ vectors living in an $m$ dimensional space, where each dimension corresponds to a row of the data matrix. The vectors' magnitude and direction indicate which rows correlate highly with which rows in the data matrix. For example, consider the distribution of points in figure~\ref{fig:covar explain}, where the coordinates of data point $d$ are governed by the $d^{\text{th}}$ row's similarity with every other row. The red points correspond to highly correlated data points arising from every $k^{\text{th}}$ row. The black points show how these rows share low information content with all other rows. For clarity, only three-dimensions are visualised, but this idea is readily extended to include all $k$ rows. Then, there exists a correlation matrix vector which has large component magnitudes in the direction of all $k^{\text{th}}$ dimensions, and small component magnitudes in all other directions. Further, the resultant matrix is symmetric and therefore has orthogonal eigenvectors, which form a basis in which data are only scaled by corresponding real eigenvalues. Because of this property, and when eigenvalues are sorted in descending order where $\{\lambda_{1}>\lambda_{2}>\ldots>\lambda_{m}\}$, the corresponding eigenvectors identify directions of maximal variance, with every subsequent eigenvector capturing the maximal variance in a direction orthogonal to all previous eigenvectors. In this sense, the eigenvectors capture the principal components of the data.

Similarly, the matrix $\bm{\mathsf{V}}$ is calculated as the eigenvectors of the matrix $\bm{\mathsf{X^\ast}}\bm{\mathsf{X}}\in\mathbb{C}^{n\times n}$. All of the previous intuition holds but this time the correlation matrix generated captures the similarity between columns of the data matrix.

Finally, the matrix $\bm{\mathsf{\Sigma}}$ holds the singular values of the matrix, which are the square roots of the non-zero eigenvalues of both $\bm{\mathsf{X}}\bm{\mathsf{X^\ast}}$ and $\bm{\mathsf{X^\ast}}\bm{\mathsf{X}}$, where the square root arises from the repeated multiplication of the data matrix $\bm{\mathsf{X}}$ to form the correlation matrix. As the singular values are related to the eigenvalues of the correlation matrix, which is a measure of the variance captured by the corresponding eigenvectors, the singular values capture the relative modal energy content of the system. Since $\bm{\mathsf{\Sigma}}$ can be calculated with knowledge of only one of $\bm{\mathsf{U}}$ or $\bm{\mathsf{V}}$, only one eigendecomposition is necessary since the remaining matrix can be calculated through a rearrangement of \eqref{PODsvd}. Typically, the eigendecomposition of the smaller correlation matrix is performed as it is often intractable to compute for the larger. While convenient, this forgoes the intuition that $\bm{\mathsf{U}}$ and $\bm{\mathsf{V}}$ are in fact both correlation-dependent matrices and whose utility depends on the structure of the data matrix $\bm{\mathsf{X}}$.

More concretely, let us define our data matrix as
\begin{equation}
\bm{\mathsf{X}}=\ \left[\begin{matrix}|&|&\ &|\\\mathbf{x}_1&\mathbf{x}_2&\cdots&\mathbf{x}_{n}\\|&|&\ &|\\\end{matrix}\right]\label{dataMatrix}
\end{equation}
which is made up of $n$ vectors, each of which belong to $\mathbb{C}^{m}$. We are interested in video data which is made up of a collection of video frames of equal time-spacing. As is typical in modal decomposition studies, the spatial measurements at a given time are all stacked into a column vector of identical size with arbitrary order but whose order is consistent across all samples. In our case, video frame $j$ is flattened into a column and forms column $j$, $\mathbf{x}_j$, of our data matrix, where each frame has $m$ pixels. 

In this format, each row represents the measured time history of a given spatial location and each column represents the spatial distribution of measurements at a given time step. Performing the SVD on $\bm{\mathsf{X}}$ therefore produces matrices $\bm{\mathsf{U}}$ and $\bm{\mathsf{V}}$ which capture the spatial and temporal correlative structure of the data, respectively. Here, the spatial measurements are pixel intensity values but could be any variable of interest, such as vorticity, velocity, or pressure. The matrix $\bm{\mathsf{U}}$ therefore captures the spatial structures which are most dominant in the data, hence why it is often termed the POD modes, while $\bm{\mathsf{V}}$ captures the time dynamics, or temporal coefficients, of the corresponding modes. To be clear, the matrix $\bm{\mathsf{U}}$ is \textit{only} termed the POD modes because of the format of the data matrix given in \eqref{dataMatrix}. Further, the spatial- and temporal-correlation matrices are decoupled and only spatial-coherence is captured by the POD modes. For this reason, this algorithm is also called space-only POD.

A number of important points need to be made here. First, discussion on the SVD highlighted how the columns of $\bm{\mathsf{U}}$, the POD modes, capture the orthogonal directions of maximal variance which can be physically interpreted as the energy of the system. Thus, the first POD mode is the spatial structure which optimally captures the entire system energy, in an $L_2$ sense, with one vector. The second mode, together with the first mode, then optimally captures the entire system energy with two orthogonal vectors, and so on. There is then nothing in this formulation which attempts to decouple independent or overlapping processes; rather, any coupling present in a system is likely to be mixed together in the POD modes.

Secondly, POD is not dependent on data ordering. The calculated eigenvectors of the correlation matrices will always identify the spatial measurements which are highly correlated. Clearly, this is not an ideal feature of a dynamical-analysis tool, but it does mean that time-sampling does not need to be uniform. The time dynamic information, captured by $\bm{\mathsf{V}}$, is, however, highly dependent on the coupling of system processes and on the time-sampling procedure.

\subsection{Understanding DMD}\label{DMDintuition}

Dynamic mode decomposition is a comparatively recent technique which has arisen from the desire to learn the physical structures and behaviours of dynamical systems directly from measured data \cite{schmid2010dynamic}. This opposes the traditional approach in engineering whereby models are produced based on analytic theory derived from first principles, which can then be used for prediction and analysis. Instead, DMD aims to capture data, infer models based off the data, and then give physical insight into processes which currently have no analytic theory available. DMD approximates the system of interest as being linear in a high-dimensional space, attempting to capture the non-linear dynamics. As this is generally not the case, concern arises over the extent to which DMD is considered valid when applied to non-linear systems. Further discussion on this is found in \S\ref{DMDconnection}.

Generally, DMD is the procedure for performing the eigendecomposition of the linear operator which maps the two matrices
\begin{equation}
	\bm{\mathsf{X}}=\ \left[\begin{matrix}|&|&\ &|\\\mathbf{x}_i&\mathbf{x}_j&\cdots&\mathbf{x}_{k}\\|&|&\ &|\\\end{matrix}\right]
\end{equation}
and
\begin{equation}
\bm{\mathsf{X}}^{\prime}=\ \left[\begin{matrix}|&|&\ &|\\\mathbf{x}_{i+1}&\mathbf{x}_{j+1}&\cdots&\mathbf{x}_{k+1}\\|&|&\ &|\\\end{matrix}\right]
\end{equation}
where the indices $i$, $j$, and $k$ are arbitrary time-steps but highlight that the only ordering that matters is that corresponding columns in $\bm{\mathsf{X}}$ and $\bm{\mathsf{X}}^{\prime}$ are shifted by one time-step. For video data, similar to the data matrix for POD, the columns hold the spatial measurements at different time-steps. In contrast to \eqref{dataMatrix}, for DMD $\bm{\mathsf{X}}$ can be written $\{\mathbf{x}_k\}^{n-1}_{k=1}$ and the time-shifted data matrix, $\bm{\mathsf{X}}^{\prime}$, can be written $\{\mathbf{x}_k\}^{n}_{k=2}$, so $\bm{\mathsf{X}},\bm{\mathsf{X}}^{\prime}\in\mathbb{C}^{m\times n-1}$. The linear operator, $\bm{\mathsf{A}}\in\mathbb{C}^{m\times m}$, then maps the columns of $\bm{\mathsf{X}}$ to the corresponding columns of $\bm{\mathsf{X}}^{\prime}$,
\begin{equation}
    \bm{\mathsf{A}}\bm{\mathsf{X}}=\bm{\mathsf{X}}^{\prime},
\end{equation}
which can be solved through
\begin{equation}
    \bm{\mathsf{A}}=\bm{\mathsf{X}}^{\prime}\bm{\mathsf{X}}^{\dagger}\label{pseudo}
\end{equation}
where the superscript $\dagger$ denotes the Moore-Penrose pseudoinverse. Typically, this is intractable as $m$ can be on the order of $10^5$ or higher from the large number of spatial measurements to be recorded. For video data, each frame can easily exceed \SI{1}{megapixel}. We are instead interested in a low-dimensional approximation of $\bm{\mathsf{A}}$ from which we can extract the behavioural properties of the linear system.

DMD builds on the foundations of POD, leveraging the utility of capturing a large percentage of the system's energy with few POD modes. To reduce the dimensionality of the approximation of $\bm{\mathsf{A}}$, DMD truncates the number of modes used for all of the analysis. Recall that the SVD of $\bm{\mathsf{X}}$ is
\begin{equation}
\bm{\mathsf{X}}=\bm{\mathsf{U}}\bm{\mathsf{\Sigma}}\bm{\mathsf{V}}^\ast\label{DMDsvd}
\end{equation}
with, this time, $\bm{\mathsf{U}}\in\mathbb{C}^{m\times q}$, $\bm{\mathsf{\Sigma}}\in\mathbb{C}^{q\times q}$, and $\bm{\mathsf{V}}\in\mathbb{C}^{n\times q}$, where $q$ is the number of modes kept to produce a $q\times q$ approximation of $\bm{\mathsf{A}}$. Prior to mode truncation, \eqref{pseudo} can be put in the form
\begin{equation}
\bm{\mathsf{A}}=\bm{\mathsf{X}}^{\prime}\bm{\mathsf{V}}\bm{\mathsf{\Sigma}}^{-1}\bm{\mathsf{U}}^\ast
\end{equation}
simply by using the definition of the SVD in \eqref{DMDsvd}, which can then be projected onto the column space of $\bm{\mathsf{U}}^\ast$ to give
\begin{equation}
    \bm{\mathsf{\tilde{A}}}=\bm{\mathsf{U}}^\ast\bm{\mathsf{A}}\bm{\mathsf{U}}
\end{equation}
where $\bm{\mathsf{\tilde{A}}}$ is the $q\times q$ approximation of $\bm{\mathsf{A}}$ which can tractably be analysed. This dimensionality reduction step can be very significant where a $10^6$-dimensional space can be embedded in a, potentially, $20$-dimensional space with low information loss. In the high-dimensional space for video data, each dimension of $\bm{\mathsf{A}}$ assumes that every pixel is independent and evolves uniquely with time. The projection onto the POD modes removes this independence assumption by defining spatial groupings of pixels whose behaviours have similar coherence in time. $\bm{\mathsf{\tilde{A}}}$ therefore captures the dynamics of the system with the POD modes forming the basis of the transformation.

An eigendecomposition of $\bm{\mathsf{\tilde{A}}}$ can now be calculated through
\begin{equation}
    \bm{\mathsf{\tilde{A}W}}=\bm{\mathsf{W}}\bm{\mathsf{\Lambda}}
\end{equation}
where $\bm{\mathsf{W}}$ and $\bm{\mathsf{\Lambda}}$ are the eigenvectors and eigenvalues of $\bm{\mathsf{\tilde{A}}}$, respectively. Knowing that this eigendecomposition was performed in a lower-dimensional representation of $\bm{\mathsf{A}}$ through the POD modes, the eigenvectors now identify contributions of the POD modes which result in oscillatory behaviour with exponential growth or decay. Therefore, when the column space of $\bm{\mathsf{X}}$ and $\bm{\mathsf{X}}^{\prime}$ are approximately equal \cite{brunton2019data}, the DMD modes can be defined through
\begin{equation}
    \bm{\mathsf{\Psi}}=\bm{\mathsf{U}}\bm{\mathsf{W}}
\end{equation}
where the columns of $\bm{\mathsf{\Psi}}$ represent the DMD modes which can be considered a linear combination of the POD modes, $\bm{\mathsf{U}}$, whose coefficients are given by the eigenvectors, $\bm{\mathsf{W}}$. Intuitively, the DMD modes use the energetically-optimal POD modes to form a basis for the approximation of $\bm{\mathsf{A}}$ whose eigenvectors identify contributions to the linear dynamics in the lower-dimensional basis.

Although DMD utilises the orthogonal modal decomposition of POD, the DMD modes themselves are generally not orthogonal \cite{schmid2010dynamic}. For practical considerations, the eigenvalues of $\bm{\mathsf{\tilde{A}}}$ are distinct, which indicates that DMD is able to decouple spatial behaviours which oscillate at different frequencies. The frequencies and exponential growth or decay rates of the DMD modes can easily be given by the corresponding phase and magnitude of the eigenvalues, respectively. Unfortunately, the convenient energy hierarchy afforded by POD is no longer available for DMD and, therefore, ranking the DMD modes by energy content can be difficult. In this work, the modal energy ranking suggested by Tu \textit{et al.} \cite{tu2014dynamic} is used. In this formulation, the modal norm, given by the $L_2$ norm, for mode $j$, $\left\lVert \mathrm{{\psi}_j}\right\rVert$, is scaled by its corresponding eigenvalue to the $n-1^{\text{th}}$ power, such that
\begin{equation}
    p_j=\lambda_j^{n-1}\left\lVert \mathrm{{\psi}_j}\right\rVert\label{DMDenergy}
\end{equation}
where $p_j$ is the energy content of the $j^{\text{th}}$ DMD mode. The scaling by the eigenvalue assumes that quickly decaying eigenvectors in the system have reduced coherence which are not necessarily representative of the system as a whole. We note that this is just one option for ordering the relative dominance of DMD modes. For example, Higham \textit{et al.} \cite{higham2018implications} demonstrates the utility of using the Fourier transform of the POD temporal coefficients, creating a power spectral density (PSD), to guide the selection of dominant DMD modes.

\subsection{Connection with Koopman operator theory and Fourier analysis}\label{DMDconnection}

DMD has close ties with Koopman operator theory and can be considered an extension of POD with Fourier analysis. In fact, under certain conditions, DMD can exactly extract Koopman eigenfunctions or can exactly compute a discrete Fourier transform (DFT).

Koopman operator theory, in essence, is a variable transformation which converts a finite-dimensional non-linear dynamical system into a possibly infinite-dimensional linear dynamical system. This transformation into a linear system is not restrictive; non-linear dynamics are still captured through the linear system, only the variables governing the system have changed. Intuitively, Koopman operator theory says that there are a set of system observables, which are a function of measurable variables, which behave linearly. The problem then becomes finding the observables. For example, consider a discrete-time system of a spatial distribution of variables $\mathbf{x}$, such as velocity or pixel-values, then the dynamical system can be written
\begin{equation}
    \mathbf{x}_{k+1}=\mathbf{f}(\mathbf{x}_{k})
\end{equation}
where $\mathbf{f}$ is some possibly non-linear function which maps from states at time-step $k$ to $k+1$. The Koopman operator, $\mathcal{U}$, acts on a set of observables, $\mathbf{g}(\mathbf{x})$, where $\mathbf{g}$ is a scalar function. This can be written
\begin{equation}
    \mathbf{g}(\mathbf{x}_{k+1})=\mathcal{U}\mathbf{g}(\mathbf{x}_{k})\label{Koopman}
\end{equation}
which is now amenable to linear analysis, where the eigenfunctions of $\mathcal{U}$, the Koopman eigenfunctions, are the spatio-temporal coherent structures of the system. DMD is an approximation of these Koopman eigenfunctions, which is exact when the scalar function $\mathbf{g}$ is just the identity function. DMD, then, can be considered an alternative to the Arnoldi method for calculating the eigendecomposition of a linear operator.

DMD also has a close relation to the discrete Fourier transform (DFT), beyond its ability to extract oscillatory bahaviour. Chen \textit{et al.} \cite{chen2012variants} showed that when data are first mean-subtracted, the DMD reduces exactly to the DFT. Comparing with the previous discussion on DMD, this is clearly restrictive as in its fullest generality, DMD is able to extract oscillatory dynamics which can also be coupled with exponential growth or decay. Subtracting the mean of the data, while a common approach in regular principal component analysis, has the consequence of also removing important dynamical information. Mean-subtraction is therefore not encouraged for most DMD studies.

\subsection{POD and DMD variants}\label{PODDMDvariants}

Although only the standard algorithms are explored and discussed in this work, some notable extensions and variants are described here as they are directly relevant to the limitations, drawbacks, and attributes of their standard counterpart. Specifically, we make no attempt to address the challenges associated with the standard techniques but intend to discuss how a validated interpretation of their results can be established, alongside a systematic approach of applying these methods. Readers are directed to the recent reviews of POD and DMD, and similar techniques, provided by Rowley and Dawson \cite{rowley2017model} and Taira \textit{et al.} \cite{taira2017modal} for an in-depth overview of modal decomposition techniques. The onus lies on the researcher to identify the technique which is most appropriate for their dataset of interest.

A variant of POD specifically for dynamical fluid systems is the balanced POD (BPOD), first introduced by Rowley \cite{rowley2005model}, which is designed to be an approximation of balanced truncation \cite{moore1981principal} which becomes intractable on high-dimensional data. Stemming from control theory, BPOD attempts to extract modes which are both highly controllable and observable. The weakly controllable or observable modes can then be truncated. Error bounds for BPOD can be set \textit{a priori}, and this approach can overcome POD's sensitivity to the empirical data used and inner product definition. However, this technique has not seen much attention in the fluid dynamics community.

Attempts have been made to incorporate frequency information into POD creating a \textit{spectral} POD algorithm (SPOD). There are, however, at least two procedures both called SPOD which attempt this in different ways. Sieber \textit{et al.} \cite{sieber2016spectral} take the data correlation matrix, as discussed in \S\ref{PODintuition}, and apply a filter which smooths the values along the diagonals of the correlation matrix. By varying the strength of this filter, SPOD can vary between a pure POD and a pure DFT to find the data augmentation which extracts modes whose spatial structures and frequencies are easily interpretable. This approach forces structure into the raw data matrix so it becomes difficult to ensure the extracted results are truly a consequence of the measured data, and not of the applied filtering process. The other procedure, where a recent discussion on POD and SPOD is given by Towne \textit{et al.} \cite{towne2018spectral}, tries to capture space-time coherence in the data, rather than the space-only coherence captured by POD. The main change is in the definition of the inner product for the correlation matrix which incorporates a sum over the temporal extent of the acquired data, not just the spatial extent.

While not necessarily an issue with high-speed video data, Brunton \textit{et al.} \cite{brunton2013compressive} provided a DMD approach to compressive sensing where requirements of complete and uniformly sampled data are relaxed. This is important for systems where spatial measurements are affected by occlusion or the probing diagnostic is subject to non-uniform time-sampling. This technique is particularly useful in two scenarios. First, if the full system of interest is only sub-sampled, it is possible to approximate the full-state DMD through what is called \textit{compressive-sampling} DMD. Second, if information on the full-state is available, \textit{compressed} DMD may be performed where the full-state is first compressed prior to applying DMD. Also borrowing ideas from compressive sensing, Jovanovi\'c \textit{et al.} \cite{jovanovic2014sparsity} suggest using a \textit{sparsity-promoting} DMD to get the best trade-off between quality of approximation and the number of modes required for a suitable approximation. To achieve this, a regulariser term is added to the approximation cost function, which is typically the $L_2$ deviation between the original data matrix and the DMD reconstruction. By including an $L_1$ regulariser, the cost function is then weighted in favour of reconstructions using fewer modes.

\textit{Optimised} DMD \cite{askham2018variable} was developed to overcome DMD's sensitivity to noise while also having improved ability at extracting hidden dynamics. This added robustness is balanced with an increased computation cost which may see up to a $50\%$ increase. The approach rephrases DMD into an exponential data fitting problem through the variable projection method which considers all snapshots of the data at the same time. This also removes any bias from the original DMD formulation and is recommended for use by the authors over the standard DMD formulation.

While DMD is an approximation of a Koopman decomposition with a scalar function, $\mathbf{g}$, equal to the identity as in \eqref{Koopman}, Williams \textit{et al.} \cite{williams2015data} generalise this notion to a dictionary of potential scalar functions with \textit{extended} DMD (eDMD). A dictionary of potential scalar functions is defined \textit{a priori}, and the eDMD algorithm identifies which combination of functions yield the best finite-dimensional linear representation of the non-linear system. This gives DMD greater flexibility to generalise to a variety of problems but this is necessarily accompanied by a much greater computational burden. As the dictionary must be defined \textit{a priori}, there is also the problem of how to select the appropriate set of candidate functions. Li \textit{et al.} \cite{li2017extended} demonstrated how to learn a suitable dictionary using a neural network to make eDMD a fully data-driven method.

Again, we emphasise that many variants exist which may be optimal for the following test cases we investigated. However, as the standard approaches continue to be used extensively and improving understanding of these approaches may aid the understanding of algorithmic variants, a detailed study on the interpretation of canonical and practical flow systems is of great benefit.

\section{Methodology}\label{methodology}

\subsection{Test cases}

\begin{figure}
    \centering
    \begin{subfigure}[b]{0.3\textwidth}
        \centering
        \includegraphics[width=0.5\textwidth]{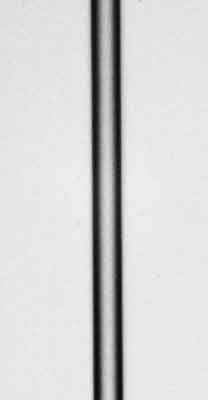}
        \caption{}
        \label{fig:laminar still}
    \end{subfigure}
    \begin{subfigure}[b]{0.3\textwidth}
        \centering
        \vspace*{0.1in}
        \includegraphics[width=0.5\textwidth]{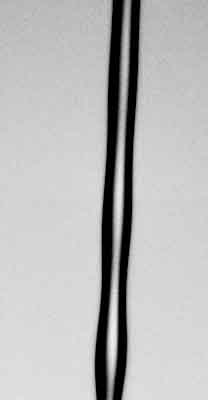}
        \caption{}
        \label{fig:dilational still}
    \end{subfigure}
    \begin{subfigure}[b]{0.3\textwidth}
        \centering
        \vspace*{0.1in}
        \includegraphics[width=0.5\textwidth]{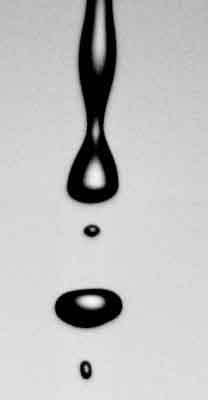}
        \caption{}
        \label{fig:breakup still}
    \end{subfigure}
    \caption{Canonical flow systems: (\subref{fig:laminar still}) laminar jet, (\subref{fig:dilational still}) dilational jet, and (\subref{fig:breakup still}) jet breakup.}
    \label{fig:canonical still}
\end{figure}

\begin{figure}
    \centering
    \begin{subfigure}[b]{0.3\textwidth}
        \centering
        \includegraphics[width=\textwidth]{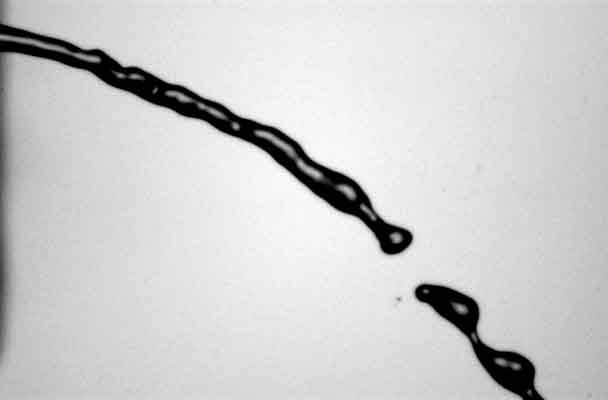}
        \caption{}
        \label{fig:lamjic still}
    \end{subfigure}
    \begin{subfigure}[b]{0.3\textwidth}
        \centering
        \vspace*{0.1in}
        \includegraphics[width=\textwidth]{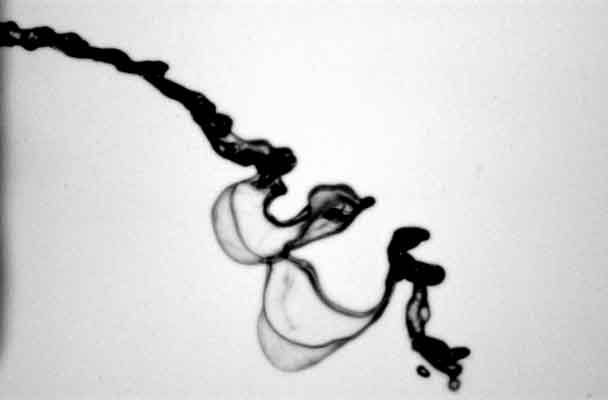}
        \caption{}
        \label{fig:bagjic still}
    \end{subfigure}
    \begin{subfigure}[b]{0.3\textwidth}
        \centering
        \vspace*{0.1in}
        \includegraphics[width=\textwidth]{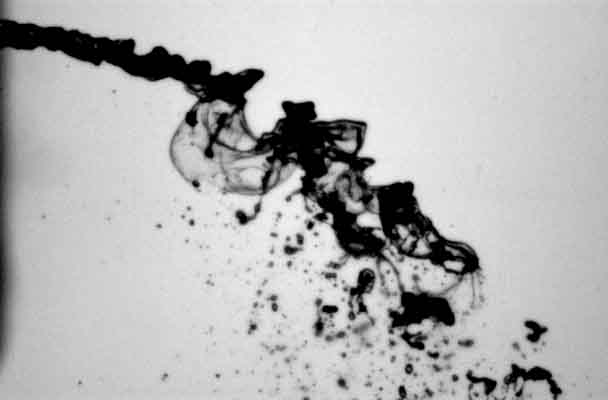}
        \caption{}
        \label{fig:multijic still}
    \end{subfigure}
    \caption{Jet in crossflow systems: (\subref{fig:lamjic still}) column breakup, (\subref{fig:bagjic still}) bag breakup, and (\subref{fig:multijic still}) multimode breakup regimes.}
    \label{fig:jic still}
\end{figure}

Several liquid injection test cases were selected, of differing complexity, to improve the understanding of how to interpret extracted modes in a hierarchical fashion. By analysing a simple system with few dominant processes, the interpretation of the modes should be straightforward and highlight how POD and DMD analyses behave. With a developed understanding of POD and DMD when applied to these simple systems, the understanding of more complex systems can be enhanced. To this end, figure~\ref{fig:canonical still} shows example snapshots of the canonical flow systems used to develop a validated understanding of POD and DMD and figure~\ref{fig:jic still} shows example snapshots of the jet in crossflow systems.

For clarity, we provide a brief qualitative description of the captured high-speed videos of the test cases. For the canonical flow systems, the liquid is injected from the top of the image and flows downstream to the bottom of the image while for all jet in crossflow cases the gaseous crossflow, which is air in this study, flows top to bottom and the liquid is injected perpendicularly to the right.

The laminar jet should provide a trivial modal decomposition; the video data show no visible oscillatory behaviour or jet perturbation. Only noise is a notable constituent of this flow. The dilational jet includes only a simple and consistent oscillation of the jet diameter which is consistent throughout the entire video. The perturbation frequency is controlled by a frequency generator and is set to 4,000~Hz. The jet breakup case is the subsequent breakup of the dilational jet where a single large droplet and an accompanying smaller satellite droplet are produced. All canonical flows are characterised by highly periodic and consistent behaviour, with the exception of the breakup location for the jet breakup case which fluctuates between upstream and downstream locations within the field of view.

Three breakup regimes were tested for the jet in crossflow systems as they have particularly distinct traits. The column breakup is the only case close to being laminar. Waves in the column are produced primarily through aerodynamic instabilities which promotes jet breakup. The bag breakup has liquid segments which first flatten out and then expand into a large bag before fully atomising. This regime, as illustrated in figure~\ref{fig:bagjic still}, has alternating segments which are undisturbed or have undergone bag breakup. The multimode breakup regime has the same characteristics as the bag breakup with the addition of the jet undergoing shear breakup. While bags still form and atomise, aerodynamic shear from the crossflow strips small droplets off the jet which may affect the jet's transition into bag breakup. The transition between these breakup regimes is governed by the liquid-to-air momentum flux ratio and the aerodynamic Weber number where here we increase both parameters to transition from column breakup to bag breakup and then from bag breakup to multimode breakup. As we are only interested in the qualitative behaviour of these flow systems we omit further details but direct readers to, for example, Wu \textit{et al.} \cite{wu1997breakup} for a more complete description of jet in crossflow breakup regimes. However, we note that we follow the \textit{effective} injection velocity formulation by Leask \textit{et al.} \cite{leask2018critical} for the momentum flux ratio calculations. For all jet in crossflow systems, both air flow and liquid injection are nominally constant but noticeable fluctuations in the jet trajectory are found in time. We denote this change in trajectory as a change in jet \textit{penetration}, where the jet penetration depth increases as the liquid penetrates further into the gaseous crossflow (i.e. further to the right in these snapshots).

In this work, we are interested in the interpretation of the results as extracted by the standard POD and DMD algorithms. As such, the application methodology is not altered specifically for each system, for example we do not optimise the numbers of frames analysed nor do we select only the most interpretable of modes. We do, however, skip discussion on modes which belong in a complex conjugate pair. These arise as an analogue to sine and cosine functions, whose properties and interpretation are identical up to a phase shift.

\subsection{Experimental setup}

A monodispersed droplet generator was used to produce the simplest canonical flow systems with well-controlled and periodic behaviours. A piezoelectric injector head with a \SI{200}{\micro m} orifice connected to a frequency generator with a 4,000~Hz signal promotes dilational behaviour in the issuing jet which eventually breaks up and forms droplets. For the jet in crossflow systems, a plain sharp-edged orifice with an orifice diameter of 0.6604~mm and an orifice length-to-diameter ratio of 4 was used to inject water into an air crossflow. The air enters a 25.4~mm $\times$ 25.4~mm square channel with the liquid injected perpendicular to the air flow. The camera used was a Vision Research Phantom v7.2 monochrome high-speed camera with an Infinity long-distance microscope lens and CFV-4 objective lens to improve the spatial resolution of the narrow spray systems. For the canonical flow systems, methanol was the test liquid, the number of analysed frames was 10,000, and the data were collected at 20,000 frames per second (fps) with a resolution of 400~$\times$~208 pixels. For the jet in crossflow systems, water was the test liquid, 5,000 frames were analysed, and data were collected at 9,708 fps with a resolution of 400~$\times$~608 pixels. For all systems, DMD used a 200-mode approximation of the linear map $\bm{\mathsf{A}}$ but, for brevity, only a subset of extracted modes are provided.

\section{Analysis of canonical flows}\label{known}

A note must first be made about the presentation of the POD and DMD modes. For all modes presented in this section and \S\ref{unknown}, no colour bars are provided. Instead, the mode colour represents the relative magnitude of the pixel values for a given mode. In all cases, blue corresponds to a large negative value, green corresponds to a value around zero, and red corresponds to large positive values. The relative pixel values capture spatial structures which have some spatio-temporal coherence associated with that structure. For DMD, the coherence is manifested in the modal frequency; the modal structure oscillates between negative and positive values at twice the corresponding modal frequency. Additionally, only the real parts of the modes are presented as this is typically done in literature but we note the interest in interpreting both real and complex modal components.

\subsection{Laminar jet}\label{known:laminar}

The laminar jet is the most trivial case for POD and DMD to give insight into. In fact, excluding noise, the video stream of the issuing laminar jet has no noticeable difference from a snapshot at any given time step. An assumption of a single dominant mode therefore makes sense, whose frequency is 0~Hz.

\begin{figure}
    \centering
    \begin{minipage}{0.7\textwidth}
    \begin{subfigure}[b]{\textwidth}
        \centering
        \includegraphics[width=\textwidth]{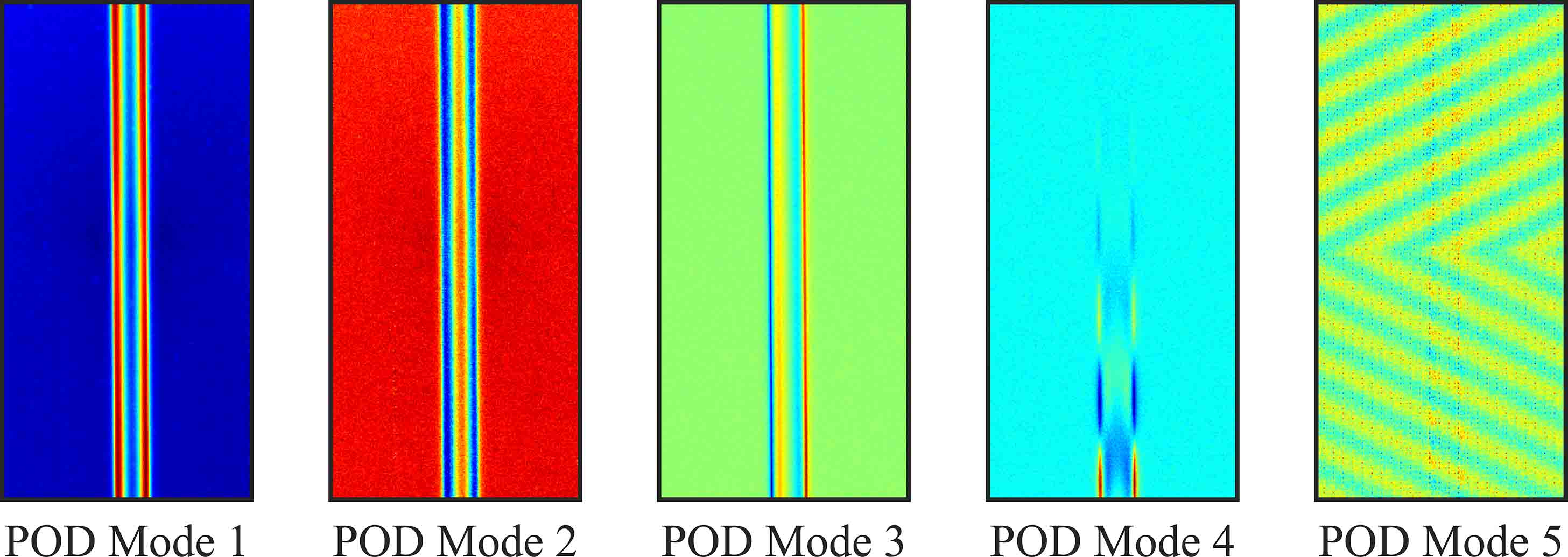}
        \caption{}
        \label{fig:lam pod modes}
    \end{subfigure}
    
    \begin{subfigure}[b]{\textwidth}
        \centering
        \vspace*{0.1in}
        \includegraphics[width=\textwidth]{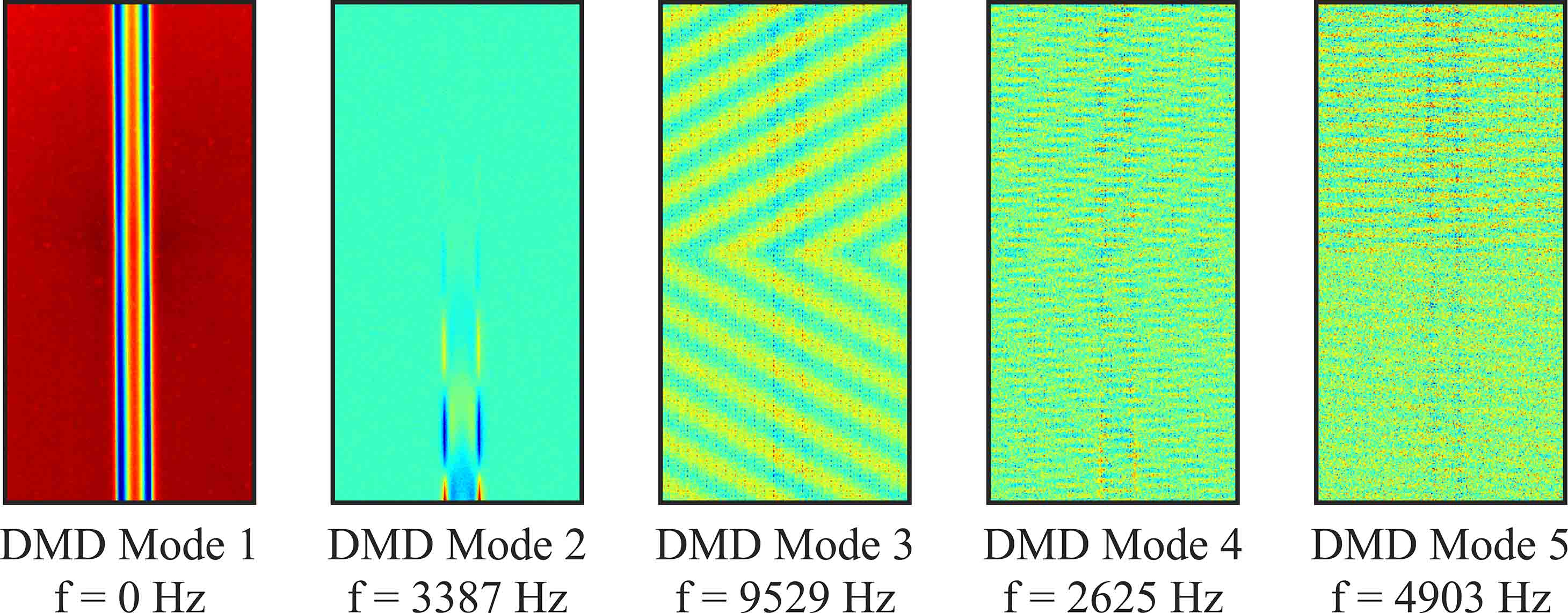}
        \caption{}
        \label{fig:lam dmd modes}
    \end{subfigure}
    \end{minipage}%
    \begin{minipage}{0.3\textwidth}
    \begin{subfigure}[b]{\textwidth}
        \centering
        \includegraphics[width=0.37794555\textwidth]{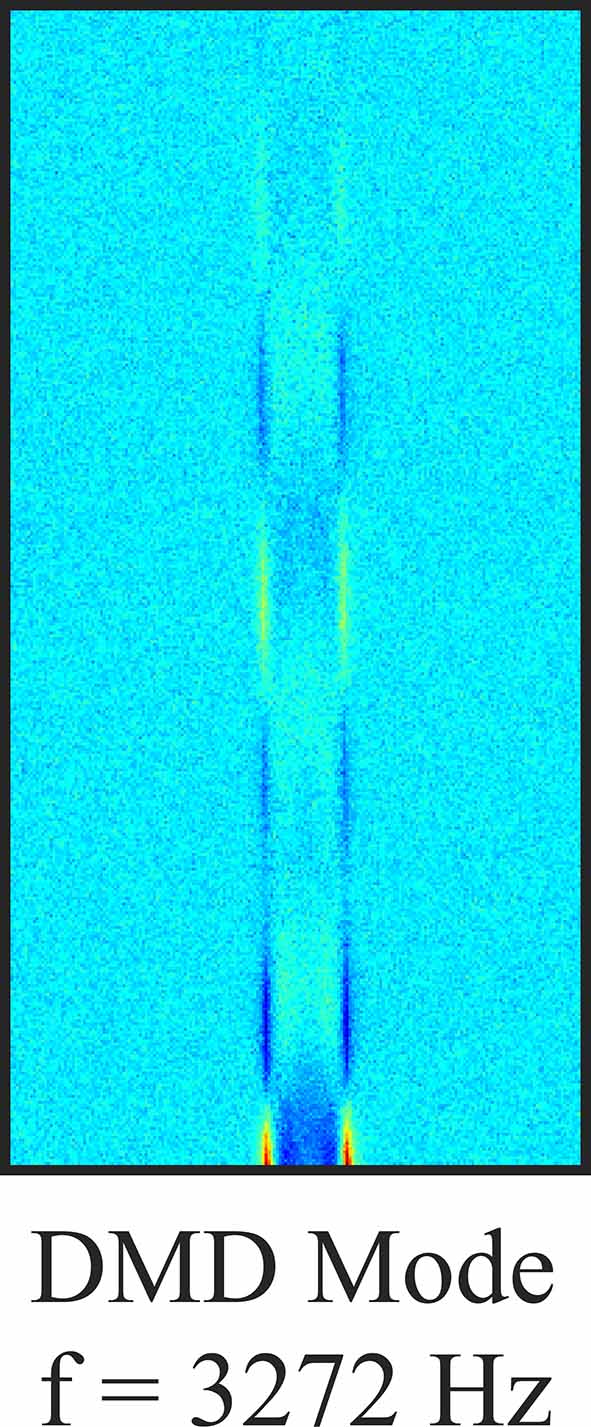}
        \caption{}
        \label{fig:lam combo}
    \end{subfigure}
    \end{minipage}
    \caption{The first five most dominant (\subref{fig:lam pod modes}) POD modes and (\subref{fig:lam dmd modes}) DMD modes for the laminar jet. (\subref{fig:lam combo}) DMD mode associated with the blue stem in figure~\ref{fig:lam dmd spec} which is not considered a unique, dominant mode.}
    \label{fig:lam modes}
\end{figure}

\begin{figure}
     \centering
     \begin{minipage}{.4\textwidth}
     \centering
     \begin{subfigure}[b]{1.5in}
         \centering
         \includegraphics[width=\textwidth]{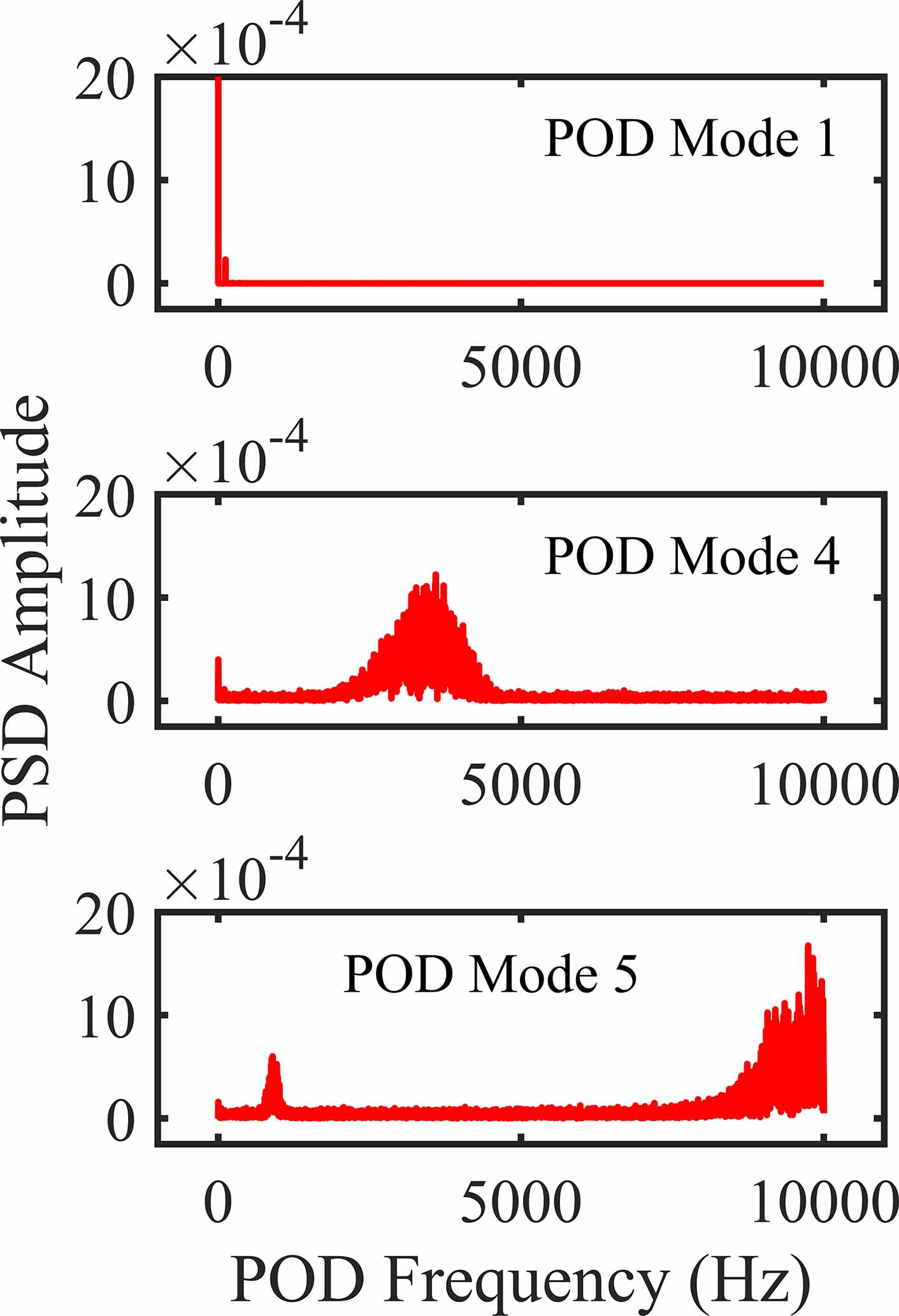}
         \caption{}
         \label{fig:lam pod spec}
     \end{subfigure}
     \end{minipage}%
     \begin{minipage}{.6\textwidth}
     \begin{subfigure}[b]{2.83in}
         \centering
         \includegraphics[width=\textwidth]{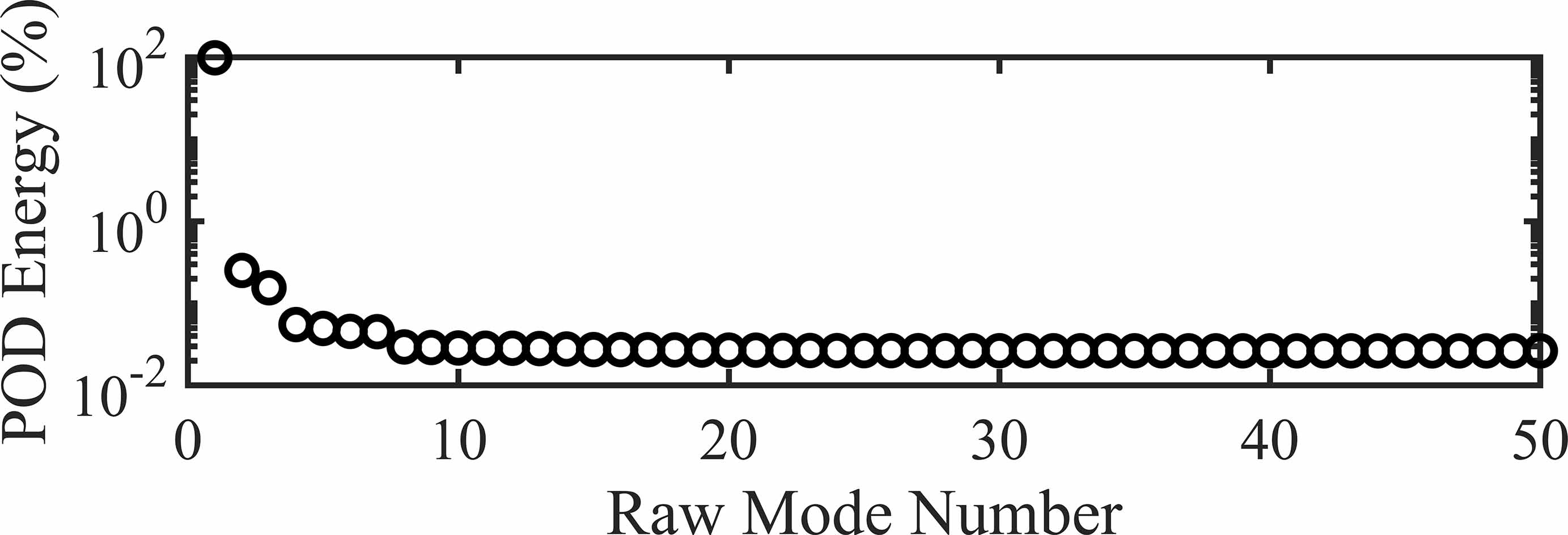}
         \caption{}
         \label{fig:lam pod dist}
     \end{subfigure}
     
     \begin{subfigure}[b]{2.8in}
         \centering
         \includegraphics[width=\textwidth]{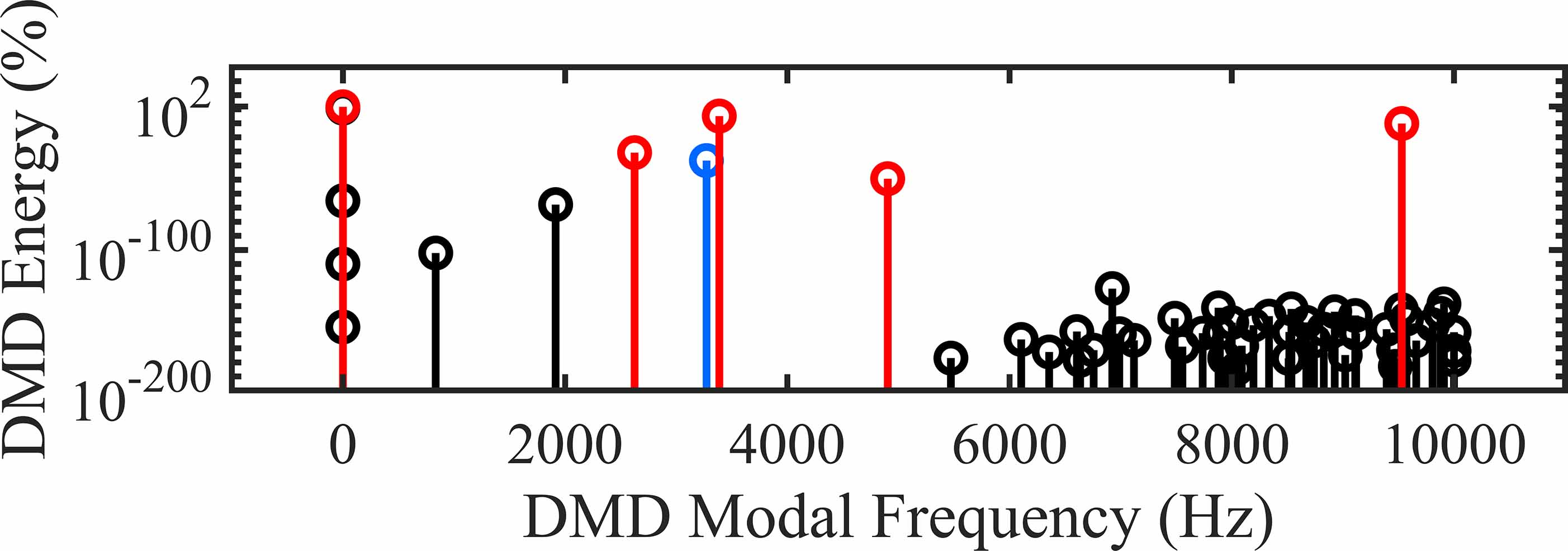}
         \caption{}
         \label{fig:lam dmd spec}
     \end{subfigure}
     \end{minipage}
        \caption{(\subref{fig:lam pod spec}) POD PSD, (\subref{fig:lam pod dist}) POD energy distribution, and (\subref{fig:lam dmd spec}) DMD spectrum for the laminar jet.}
        \label{fig:lam graphs}
\end{figure}

The first five modes for POD and DMD are given in figure~\ref{fig:lam modes} with some accompanying results in figure~\ref{fig:lam graphs}. A well known result in the use of both POD and DMD is that the first mode captures the time-averaged system, at least for systems where mean-subtraction has not been applied, and the mode oscillates at 0~Hz. We note that the inverted values of the first modes have no effect on interpretation, so the first POD and DMD modes are equivalent. While DMD is spectrally clean in that it outputs a single modal frequency per DMD mode, it is only possible to produce a power spectral density (PSD) for each POD mode, as given in figure~\ref{fig:lam pod spec}, by taking the Fourier transform of the modal temporal coefficients. For the first three POD modes, the PSD captures the expected 0~Hz frequency, but also captures non-zero frequency spikes such as at 120~Hz for the first mode. The second and third POD mode PSD graphs are not given as they are similar to that of the first mode. The fact that the first three POD modes share frequency content is not desirable as they are likely to capture similar processes, which is evidently the case when looking at their modal structures as the second mode is a noisier and inverted version of the first mode.

Due to the energy-optimality property of POD, we can extract a modal energy distribution, given in figure~\ref{fig:lam pod dist}. We denote this the \textit{raw} modal energy distribution, as it displays the relative energy content for each \textit{raw} mode extracted through POD. In some cases, these modes come in complex conjugate pairs; modes with an equivalent PSD, energy content, and whose spatial structures have a 90\degree\ phase offset (analogies can be made to sine and cosine waves). Due to this dependence, we treat the complex conjugate pairs as a single mode, and differentiate the \textit{raw} modes given in the POD energy distribution with the modes presented in, for example, figure~\ref{fig:lam pod spec} and figure~\ref{fig:lam pod modes} which are energy-ordered modes which are distinct from the rest.

Following this, the POD energy distribution finds that the first mode captures 98.2~\% of the entire system energy. This agrees with our intuition and the first POD mode evidently captures the laminar jet. According to the energy distribution, however, another two modes are more dominant than the other modes. As mentioned previously, their PSD graphs are similar to the first mode and no discernible extra information is provided by these modes. This may suggest that only the first mode is truly needed to represent the entire system. POD modes four and five clearly indicate the contrary.

DMD picks up modes similar to the fourth and fifth POD mode but they are given greater weighting in the energy distribution, where the red stems in the DMD spectrum correspond to the displayed DMD modal structures. Interestingly, the second DMD mode captures spatial structures with periodic spacing which oscillates at 3,387~Hz. The PSD of the fourth POD mode attempts to identify the oscillation frequency but provides a band of frequencies which are constituents of the mode. The peak frequency is 3,592~Hz, which agrees with DMD, but it suffers from its imprecise nature. By comparing the POD and DMD modes, we see that POD incorporates noise into this mode, which may explain the broad frequency band associated with this mode. DMD is able to decouple the noise from this modal structure. From the video of the laminar jet, there is no discernible dilational behaviour, so this would constitute an underlying \textit{hidden} mode, perhaps based on a subtle but consistent change in pixel values. As this behaviour is similar to the dilational jet investigated in \S\ref{known:dilational}, it is possible that this mode is capturing self-induced dilational behaviour. This mode is discussed more in \S\ref{known:dilational}.

The remaining DMD modes, and the fifth POD mode, still capture intuitive structures: noise associated with the sensor of the camera. These structures are mirrored about the horizontal centreline of the mode which corresponds to the location where there is a change in the camera sensor quadrant. Again, while DMD assigns a single frequency for each mode, POD assigns a frequency band for the fifth mode which has a peak at 9,741~Hz. This band may be more reasonable as it is capturing a noisy process, but a second peak is present at 900~Hz which is likely to be erroneous. While it has been found for complex flows that POD can mix processes of multiple frequencies together \cite{higham2018implications}, this same finding can be found for the most trivial of flow cases.

Due to the true low-dimensional subspace that the laminar jet can be embedded, namely through only two DMD modes, the next most pertinent modes are an artifact of the camera noise which gives confidence that the liquid system can be fully characterised in just two DMD modes. In contrast with POD, the ordering of the modes suggests that only one mode is needed to fully represent the system. This agrees with our initial intuition but it fails to consider lower energy modes which may contribute to the system in a meaningful way. Of course, the second and third POD modes can be neglected in order to consider the fourth POD mode, but this ignores the purpose of POD to extract modes in an energy-optimal way.

While POD provides an energy hierarchy for the extracted modes, the selection of the most dominant DMD modes is not as robust. As will be clear in more complex systems, although DMD produces a mode with a single accompanying frequency, modes tend to cluster around frequencies which are prominent in the system. For example, figure~\ref{fig:lam dmd spec} shows a cluster of modes all at 0~Hz with relatively large DMD energies. The first five most energetic DMD modes, according to figure~\ref{fig:lam dmd spec}, are all at 0~Hz, which are highly related to the first three POD modes. While POD indicates energy dominance \textit{per mode}, DMD indicates energy dominance \textit{per frequency}. Therefore, we choose to select dominant modes based on their DMD energy \textit{and} their modal frequency, with the intuition that all modes with a lesser energy but similar frequency derive from the combination of the most dominant mode at that frequency with either noise or other modes. As an example, the DMD mode given by the blue stem in figure~\ref{fig:lam dmd spec} has a modal structure, given in figure~\ref{fig:lam combo}, which is similar to the second DMD mode but with more noise and has a similar oscillatory frequency. As this mode appears to be dependent on at least the second DMD mode, it is not considered a distinct and dominant mode. In general, this approach works well, as all modes at 0~Hz have a combined DMD energy of over 99\%, which agrees with the energy content of the first POD mode.

The POD and DMD modes in reality do not differ greatly for this system. Most of the DMD modes have equivalent POD modes in structure, although they may appear at different energetic levels. For example, the second DMD mode is similar to the fourth POD mode. The main difference in these techniques here is \textit{how} the dominant modes are selected.

\subsection{Dilational jet}\label{known:dilational}

For the dilational jet, known system frequencies and perturbations were available and serve as guidelines for how the modes should be interpreted. The understanding gained from the laminar jet now should make interpreting the results for the dilational jet case easier. We expect the same modes for the laminar jet to exist, but this time a dominant mode capturing the 4,000~Hz perturbation is expected to be captured for both algorithms.

\begin{figure}
    \centering
    \begin{minipage}{0.7\textwidth}
    \begin{subfigure}[b]{\textwidth}
        \centering
        \includegraphics[width=\textwidth]{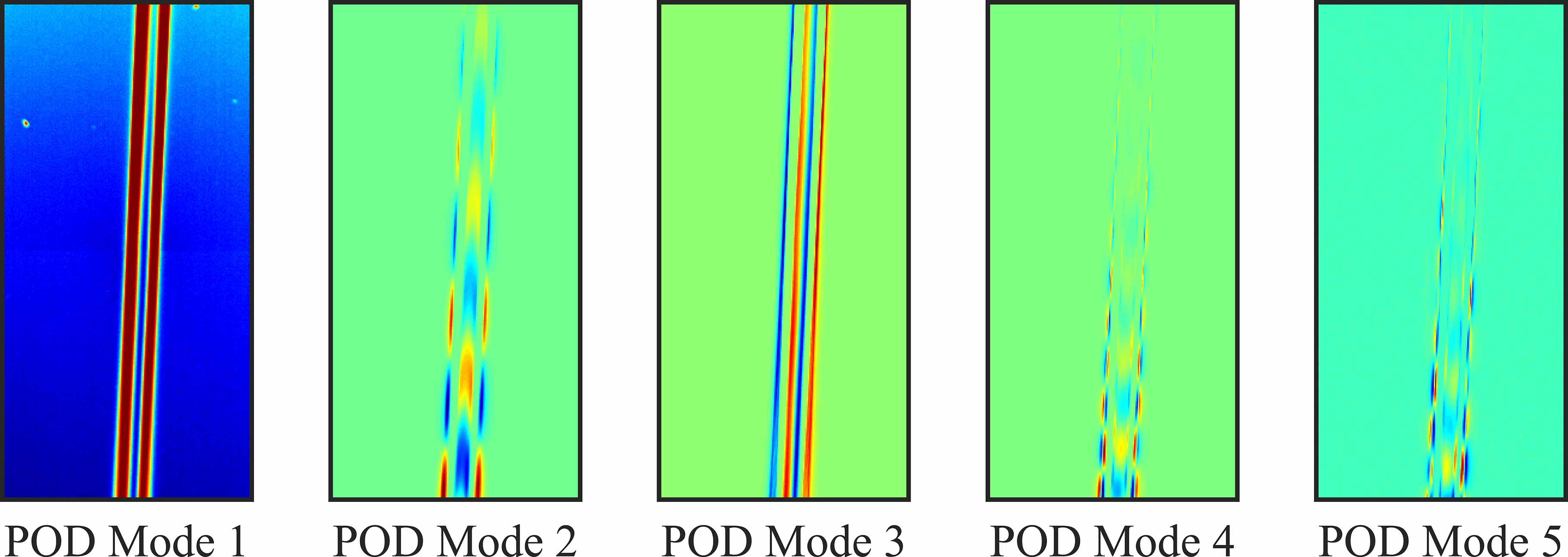}
        \caption{}
        \label{fig:sin pod modes}
    \end{subfigure}
    
    \begin{subfigure}[b]{\textwidth}
        \centering
        \vspace*{0.1in}
        \includegraphics[width=\textwidth]{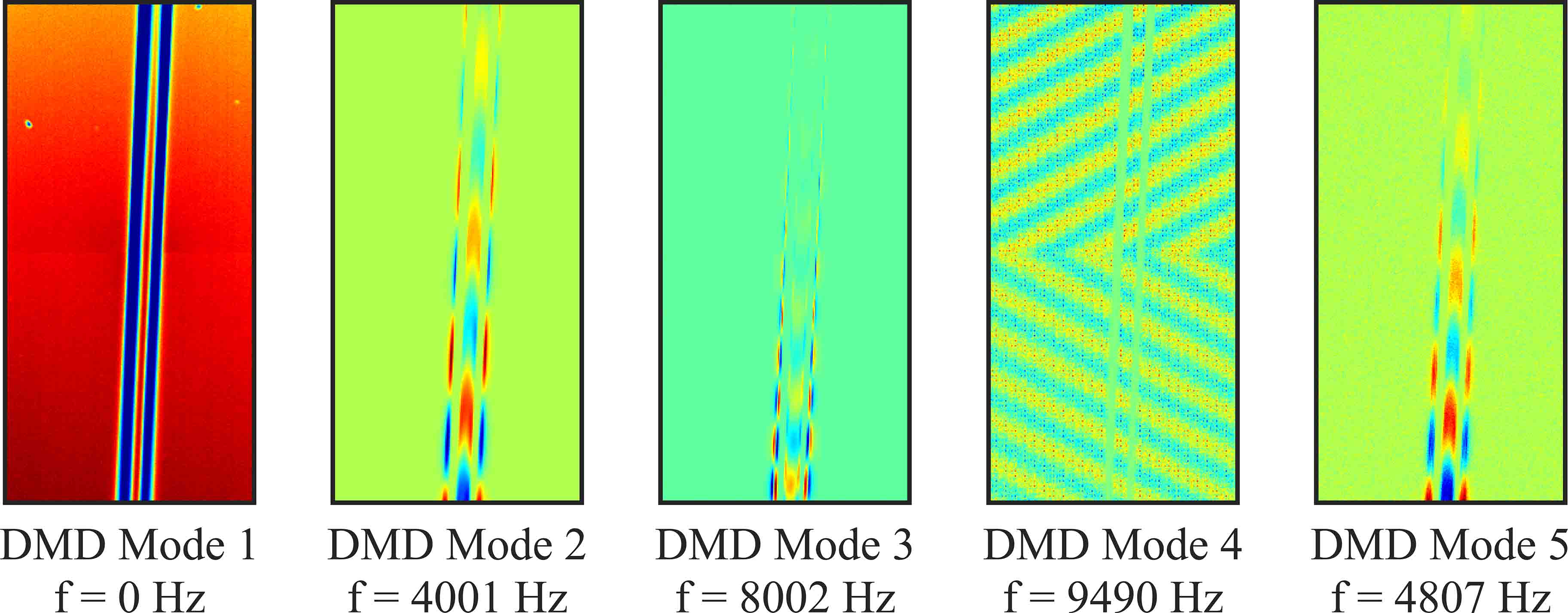}
        \caption{}
        \label{fig:sin dmd modes}
    \end{subfigure}
    \end{minipage}%
    \begin{minipage}{0.3\textwidth}
    \begin{subfigure}[b]{\textwidth}
        \centering
        \includegraphics[width=0.37794555\textwidth]{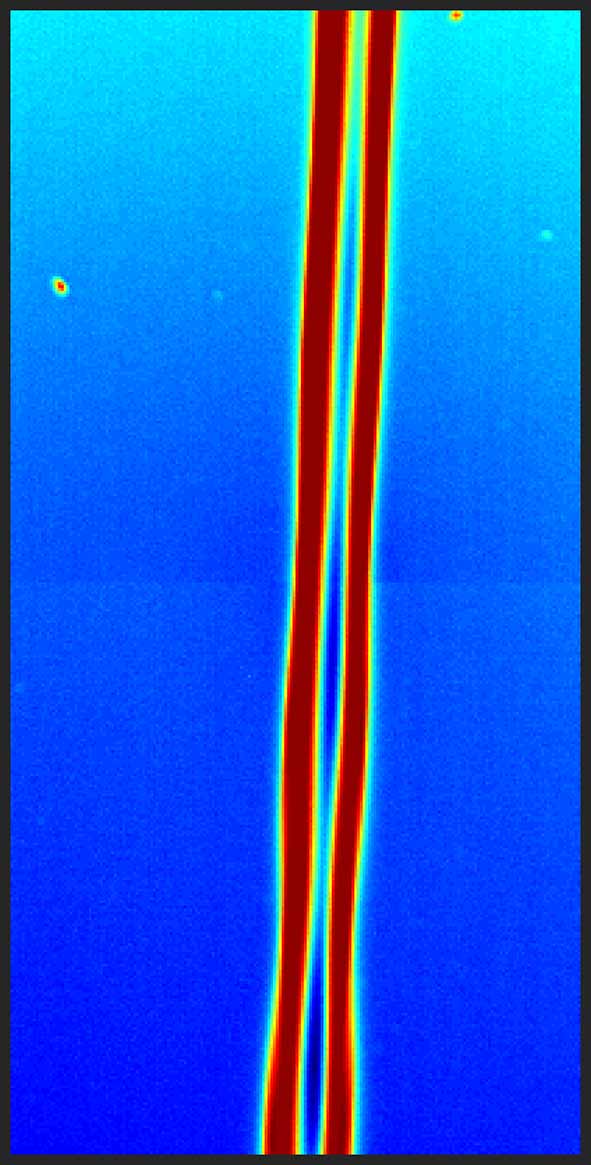}
        \caption{}
        \label{fig:sin reconstruct}
    \end{subfigure}
    \end{minipage}
    \caption{The first five most dominant (\subref{fig:sin pod modes}) POD modes and (\subref{fig:sin dmd modes}) DMD modes for the dilational jet. (\subref{fig:sin reconstruct}) System reconstruction using the first two modes from either POD or DMD.}
    \label{fig:sin modes}
\end{figure}

\begin{figure}
     \centering
     \begin{minipage}{.4\textwidth}
     \centering
     \begin{subfigure}[b]{1.5in}
         \centering
         \includegraphics[width=\textwidth]{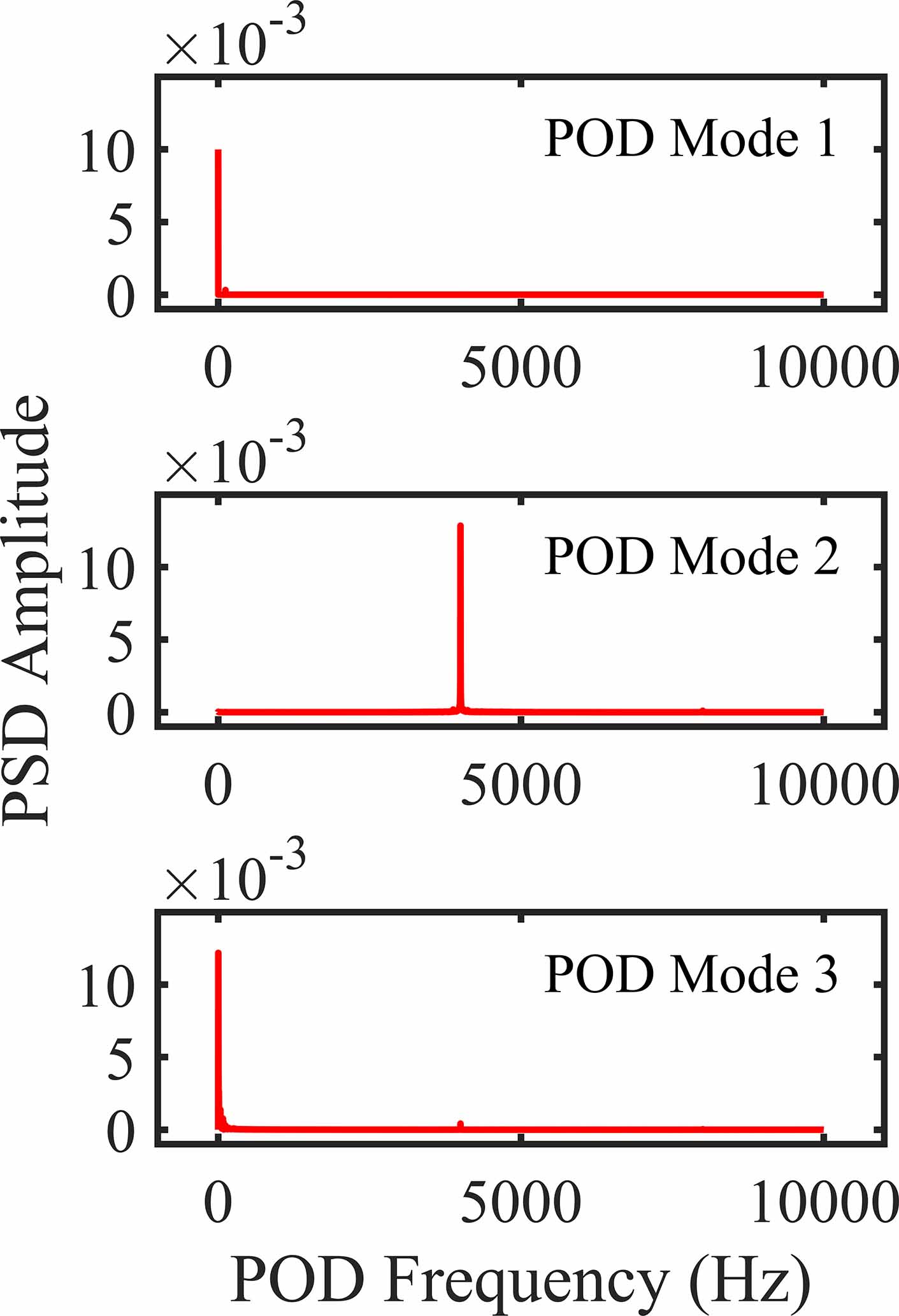}
         \caption{}
         \label{fig:sin pod spec}
     \end{subfigure}
     \end{minipage}%
     \begin{minipage}{.6\textwidth}
     \begin{subfigure}[b]{2.83in}
         \centering
         \includegraphics[width=\textwidth]{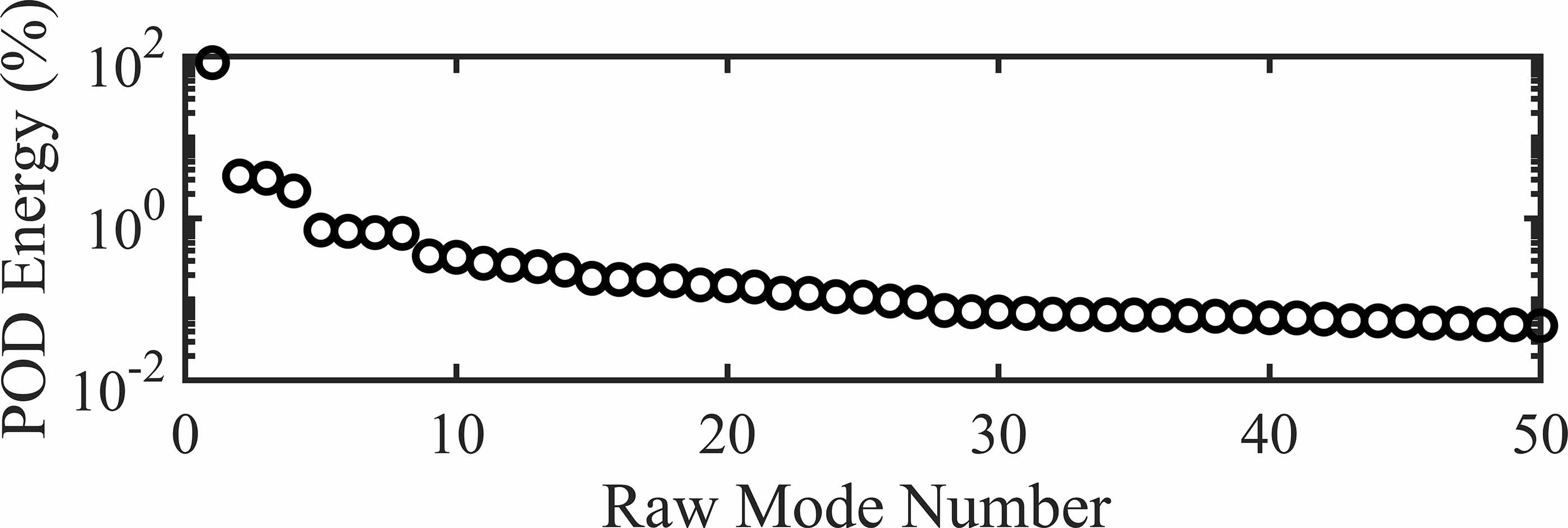}
         \caption{}
         \label{fig:sin pod dist}
     \end{subfigure}
     
     \begin{subfigure}[b]{2.8in}
         \centering
         \includegraphics[width=\textwidth]{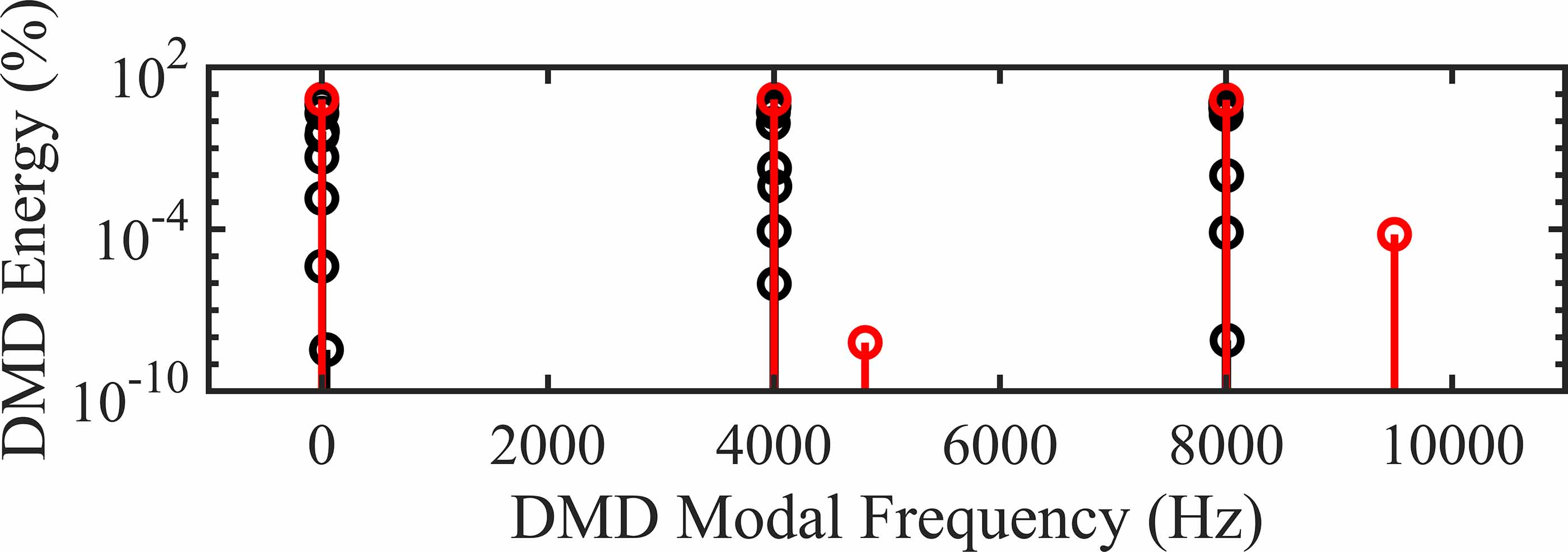}
         \caption{}
         \label{fig:sin dmd spec}
     \end{subfigure}
     \end{minipage}
        \caption{(\subref{fig:sin pod spec}) POD PSD, (\subref{fig:sin pod dist}) POD energy distribution, and (\subref{fig:sin dmd spec}) DMD spectrum for the dilational jet.}
        \label{fig:sin graphs}
\end{figure}

The first five POD and DMD modes for the dilational jet are given in figure~\ref{fig:sin pod modes} and figure~\ref{fig:sin dmd modes}, respectively, with the corresponding modal information given in figure~\ref{fig:sin graphs}. From the POD energy distribution in figure~\ref{fig:sin pod dist}, there appears to be three modes which dominate the system (where two \textit{raw} modes form a complex conjugate pair), but this dilineation is weaker than the laminar jet owing to the, albeit slight, increase in system complexity. From the laminar jet case, POD suggested one dominant mode with an energy greater than 1\%, so it makes sense for there to be more modes also contributing to the dilational jet system at a higher energy level.

As was discussed for the laminar jet, the second DMD mode and fourth POD mode appeared to capture the spatial regions which contribute to the dilational behaviour of the jet. With greater dilational behaviour being induced, now both techniques capture this behaviour in their second mode. Both POD and DMD correctly identify the perturbation frequency of 4,000~Hz, with the PSD for the POD mode in figure~\ref{fig:sin pod spec} confidently capturing a single frequency, and their spatial structures capture locations where liquid is more likely to be present (indicated by red) or less likely to be present (indicated by blue) at a given time step. By identifying either red or blue values in a mode in isolation, the general structure of a dilational jet is produced. The physical interpretation of this can be confirmed by performing an approximate system reconstruction using the first and second modes through a weighted sum of these spatial structures with their corresponding singular values, whose result is given in a figure~\ref{fig:sin reconstruct}. Note, this reconstruction works for both POD and DMD, the only difference is that the second DMD mode has a 90\degree\ phase offset to the second POD mode, which provides an example of what a complex conjugate pair looks like, and so the reconstruction would maintain this 90\degree\ phase offset.

The third POD mode identifies the edges of the jet with an accompanying PSD which has peak amplitudes between 2~Hz and 100~Hz. On reinspection of the dilational jet video, the jet exhibits irregular transverse motion, moving the jet about a pixel or two side-to-side. This third POD mode can be said to capture this motion, whose irregularity is embedded in the PSD, as the jet moves preferentially to either the blue or red values at a given moment in time. While POD picks up on this subtle motion, it does not appear in the first five dominant DMD modes.

\begin{figure}
    \centering
    \includegraphics[width=0.85\textwidth]{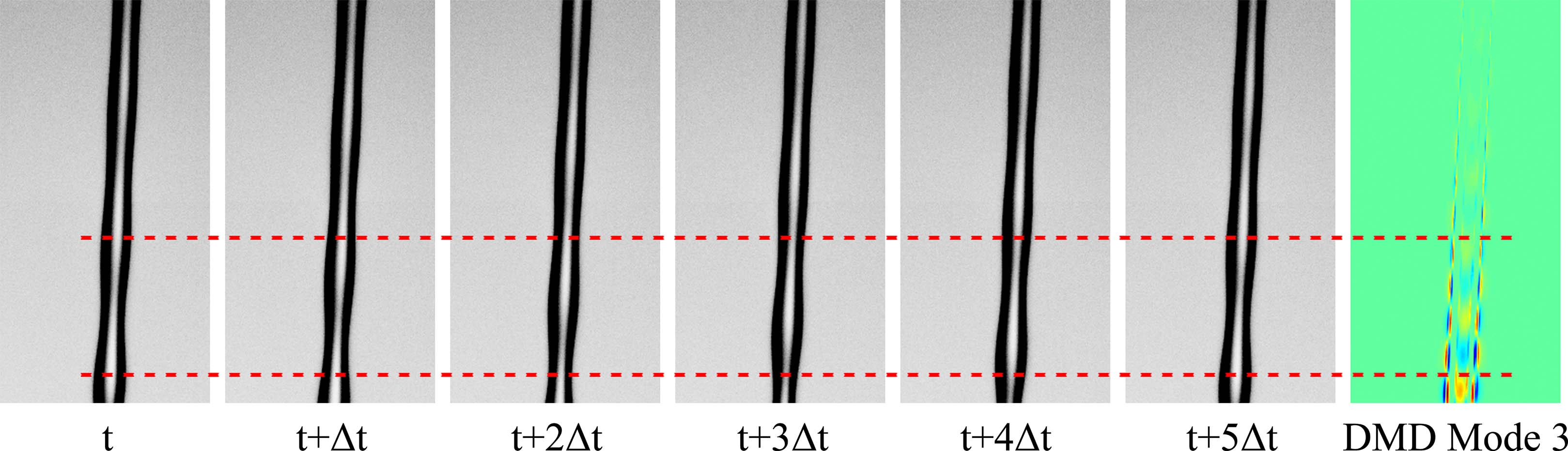}
    \caption{Six consecutive snapshots of the raw video data, the red lines highlight the five-time-step periodicity of the jet. Inherent spatial scales and motion are captured by the second DMD mode whereas the third DMD mode only captures motion at twice the frequency of the second mode.}
    \label{fig:dilational motion}
\end{figure}

The third DMD mode captures finer spatial structures at twice the frequency of the second DMD mode. This corresponds to a higher harmonic of the second mode, where the second mode can be considered a \textit{fundamental} mode, which is commonly found in systems exhibiting periodic processes (see, for example, \cite{tu2014dynamic}). These harmonics often share very similar structure as the fundamental modes, whose spatial structures depend on the order of their harmonic. In this situation, the third DMD mode is the second harmonic of the second mode whose spatial structures are a factor of two smaller. We know that an accurate reconstruction of the system can be made using only the first two modes, but adding the third mode into the reconstruction does not provide any significant changes. This is expected based on its reduced energy contribution coupled with its finer spatial scales, which do not identify the true dilational behaviour of the jet, so what does this mode mean physically? Knowing the frequency of the mode and the sampling frequency of the video data, we find that the second DMD mode correctly captures the spatial scales of interest and that the spatial structures oscillate between the positive and negative values at \textit{double} the modal frequency. Intuitively, for a given mode, if liquid is present where values are positive, liquid is then unlikely to be present where values are negative, and the modal frequency tells us how frequently liquid is present where the values are positive. The third DMD mode does not explicitly capture the spatial distribution of the liquid in the system, but rather the \textit{motion} of the liquid associated with the spatial structure of the associated fundamental mode. More quantitatively, the second DMD mode identifies the spatial structures and their periodic motion through positive (or negative) values, which occurs every $f_s/f=20000/4000=5$ time steps, where $f_s$ is the sampling frequency and $f$ is the modal frequency, while the third DMD mode identifies the periodic motion of those same spatial structures every $f_s/f=20000/8000=2.5$ time steps. This is shown in figure~\ref{fig:dilational motion} with the five-time-step periodicity of the jet, highlighted by the red lines. Note that the third DMD mode's spatial scales are not found in the raw video data. At twice the fundamental frequency, the third DMD mode captures two periods of finer spatial scale structures which only represent motion at half the fundamental temporal scale.

\begin{figure}
    \centering
    \includegraphics[width=0.7\textwidth]{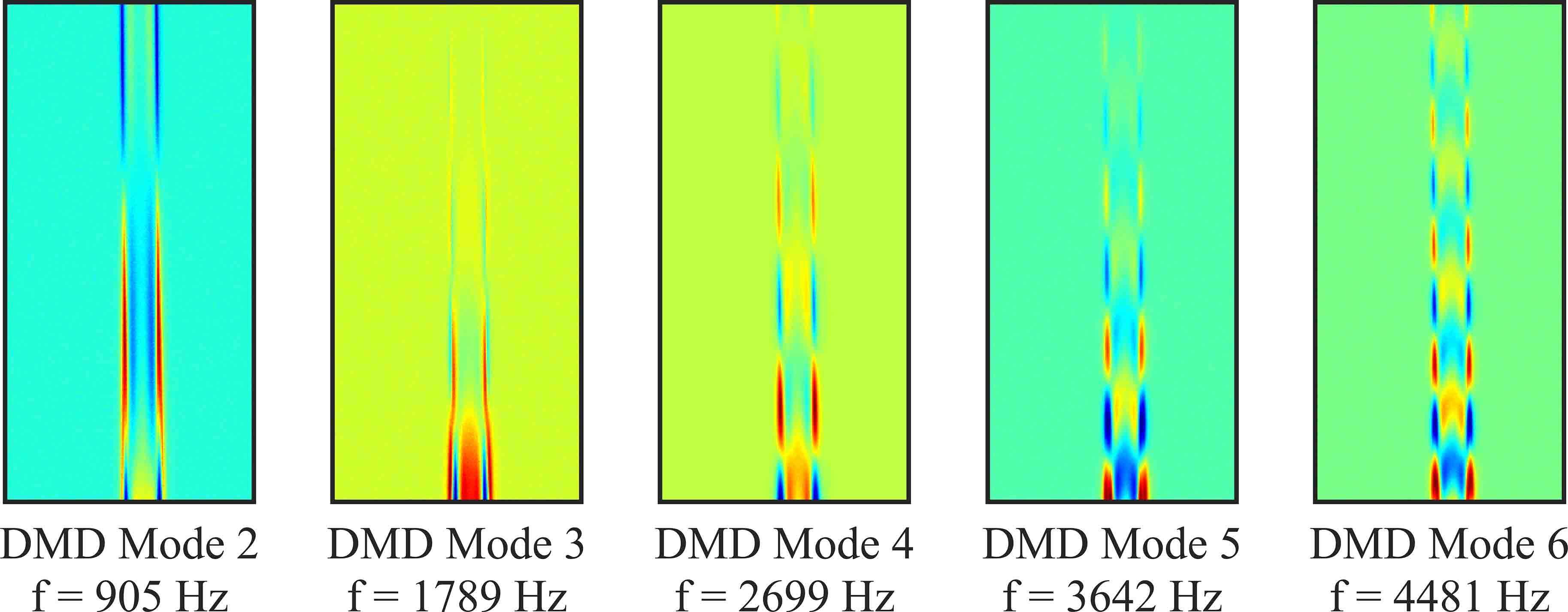}
    \caption{The first five non-zero frequency DMD modes for a dilational jet produced with a perturbation frequency of 900~Hz. Note the correlation between the higher harmonics' frequency and the decrease in spatial scales.}
    \label{fig:dilational harmonics}
\end{figure}

To further look at the role of harmonic modes through POD and DMD, a second dilational jet was investigated whose perturbation was lowered to 900~Hz to allow a number of higher harmonics to potentially be captured. Figure~\ref{fig:dilational harmonics} shows the first five non-zero frequency DMD modes for this dilational jet. The fundamental mode and frequency are captured in the second mode, which correctly identifies the spatial extent and frequency of the dilational jet. All subsequent modes are almost perfect higher harmonics of the fundamental mode, again whose spatial scales are reduced by a factor corresponding to their harmonic number. For example, the fourth DMD mode is the third harmonic which captures spatial scales a factor of three smaller than the scales of the fundamental frequency. Only the first four higher harmonics are shown for brevity, but this trend continues up until the Nyquist limit. We note again, for clarity, that the higher harmonic modes do not capture spatial waves or underlying structures, rather these modes capture the \textit{motion} of the dilational jet. These harmonics can be interpreted as identifying regions where if liquid is present, such as at maximal values, then the liquid will move to the adjacent downstream minimal values in $f_s/\left(2f\right)$ time steps. Care must then be taken to identify modes which capture pure motion content or a combination of motion and structure content.

For the case presented in figure~\ref{fig:sin modes}, the fourth POD mode resembles the third DMD mode, which demonstrates DMD's ability to extract the most dominant behaviours in the least number of modes for this case. From the DMD spectrum in figure~\ref{fig:sin dmd spec}, the system is dominated by three modes (where, again, we only take the most dominant mode for a given frequency band) but it is unclear whether any other modes are truly capturing an independent process in the system. The POD energy distribution suggests three dominant modes, as does the DMD spectrum, so it appears well founded to truncate after the third mode. However, as for the laminar jet case, the fourth DMD mode captures the sensor noise which is a true underlying process which is affecting the data. The information carried by this mode is clearly insignificant compared to the other modes, but if the purpose of POD and DMD is to extract the true underlying processes, irrespective of their relative contributions, it would be preferred if it were easy to extract these modes in a systematic way. With the current formulation, the relative energy along with a study of the modal structures is the only way to find the true processes. As we look at more complex systems, the ability to do this is greatly hampered.

The fifth DMD mode further highlights the rationale behind only selecting one mode per frequency band; the modes are often a combination of multiple true modes or modes with noise. This mode has the general structure of the second DMD mode, perhaps mixed with the third DMD mode to reduce the spatial scales captured, and in combination with noise to give a grainy background to the mode. While it is not impossible for this to be a true underlying process (e.g. the jet perturbation frequency may have produced a spectrum of disturbances but at much lower amplitudes), few modes are present around this frequency and the mode itself has not separated distinct spatial structures from noise. It is unlikely that a true underlying mode has a combination of both distinct spatial structures and noise.

\subsection{Jet breakup}\label{jet breakup}

The analysis of the breakup of a liquid jet is of particular interest as it has many similarities with both primary and secondary atomisation in practical systems. Similar to the dilational jet, a perturbation frequency of 4,000~Hz was applied which produces large droplets and accompanying satellite droplets, both of which are expected to be captured through modal analysis.

\begin{figure}
    \centering
    \begin{subfigure}[b]{0.7\textwidth}
        \centering
        \includegraphics[width=\textwidth]{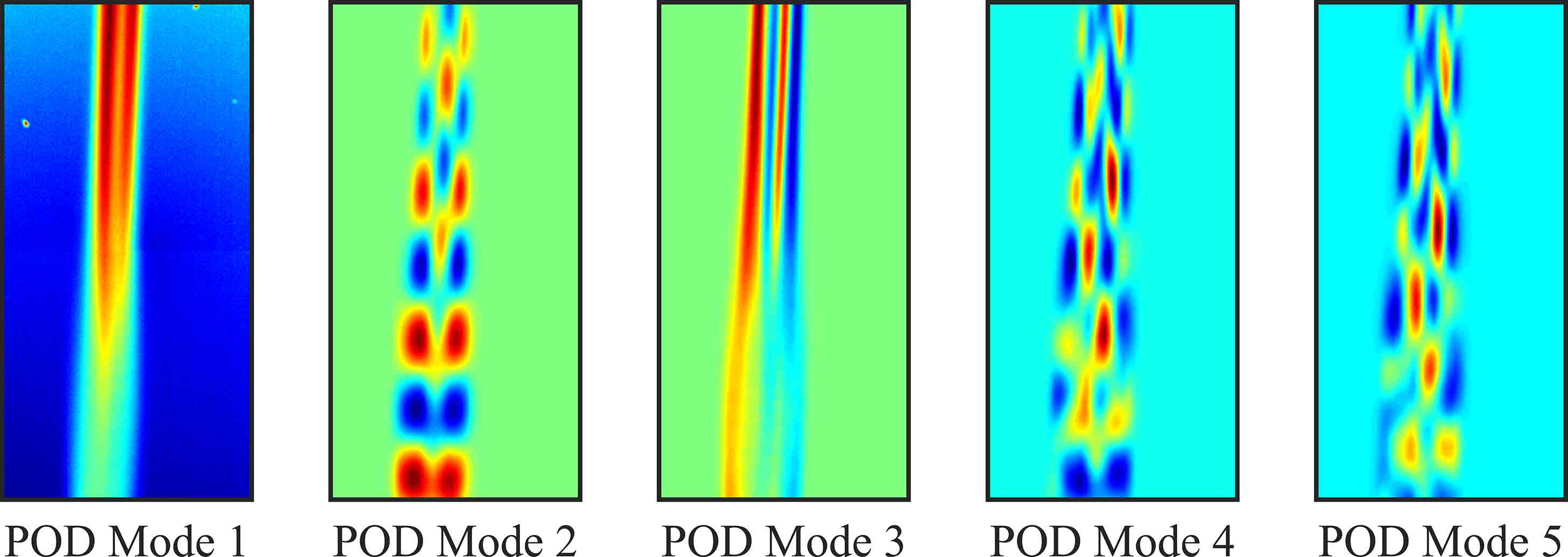}
        \caption{}
        \label{fig:break pod modes}
    \end{subfigure}
    
    \begin{subfigure}[b]{0.7\textwidth}
        \centering
        \vspace*{0.1in}
        \includegraphics[width=\textwidth]{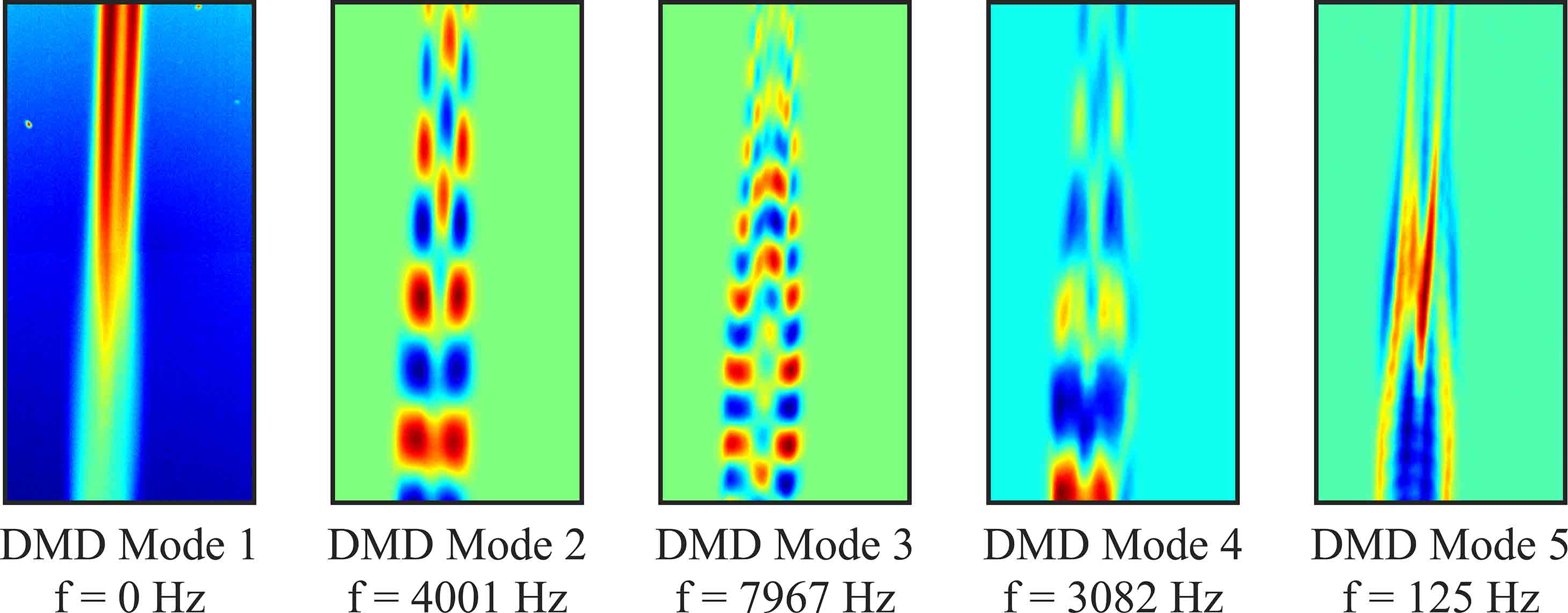}
        \caption{}
        \label{fig:break dmd modes}
    \end{subfigure}
    \caption{The first five most dominant (\subref{fig:break pod modes}) POD modes and (\subref{fig:break dmd modes}) DMD modes for the jet breakup case.}
    \label{fig:break modes}
\end{figure}

\begin{figure}
     \centering
     \begin{minipage}{.4\textwidth}
     \centering
     \begin{subfigure}[b]{1.5in}
         \centering
         \includegraphics[width=\textwidth]{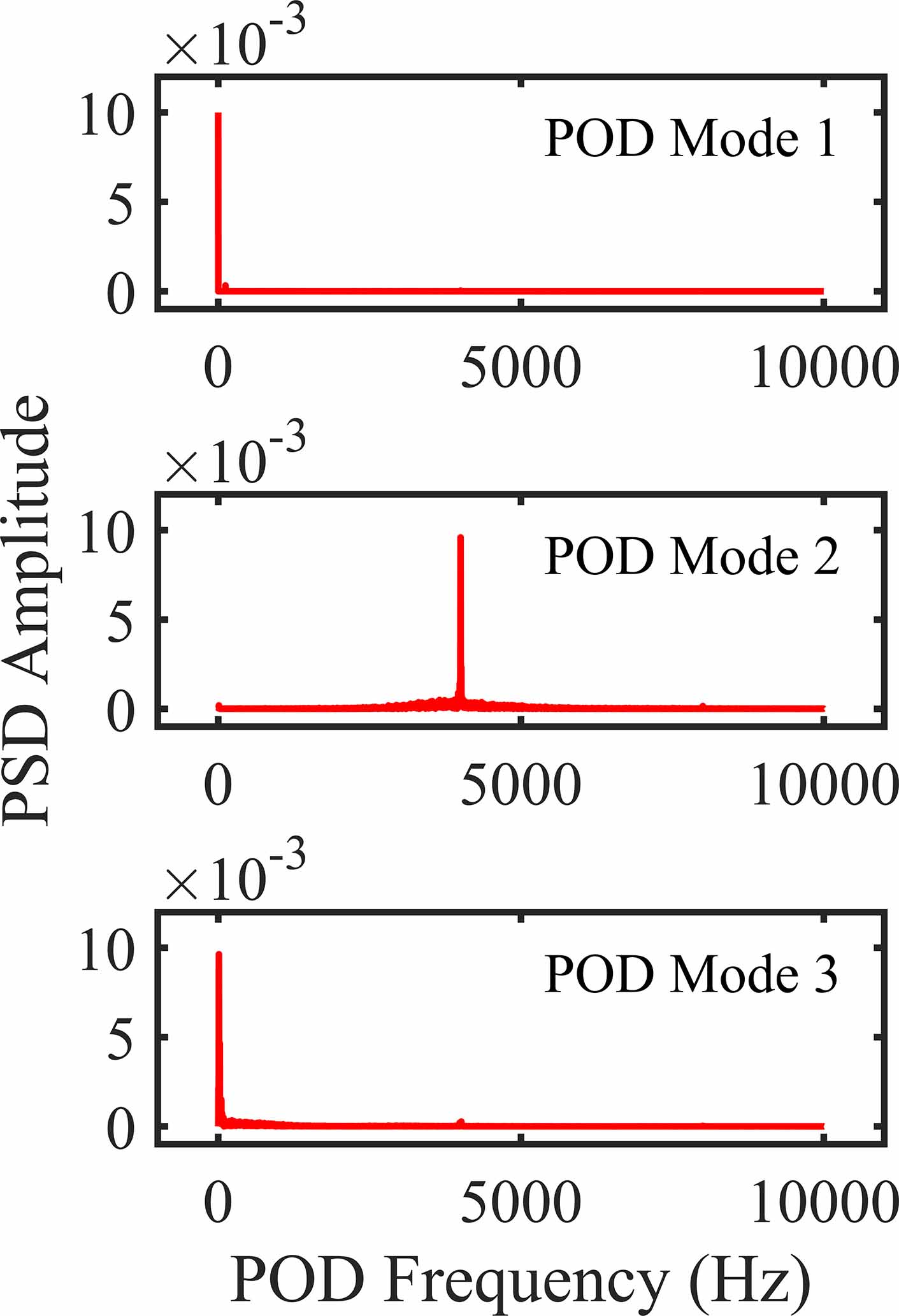}
         \caption{}
         \label{fig:break pod spec}
     \end{subfigure}
     \end{minipage}%
     \begin{minipage}{.6\textwidth}
     \begin{subfigure}[b]{2.83in}
         \centering
         \includegraphics[width=\textwidth]{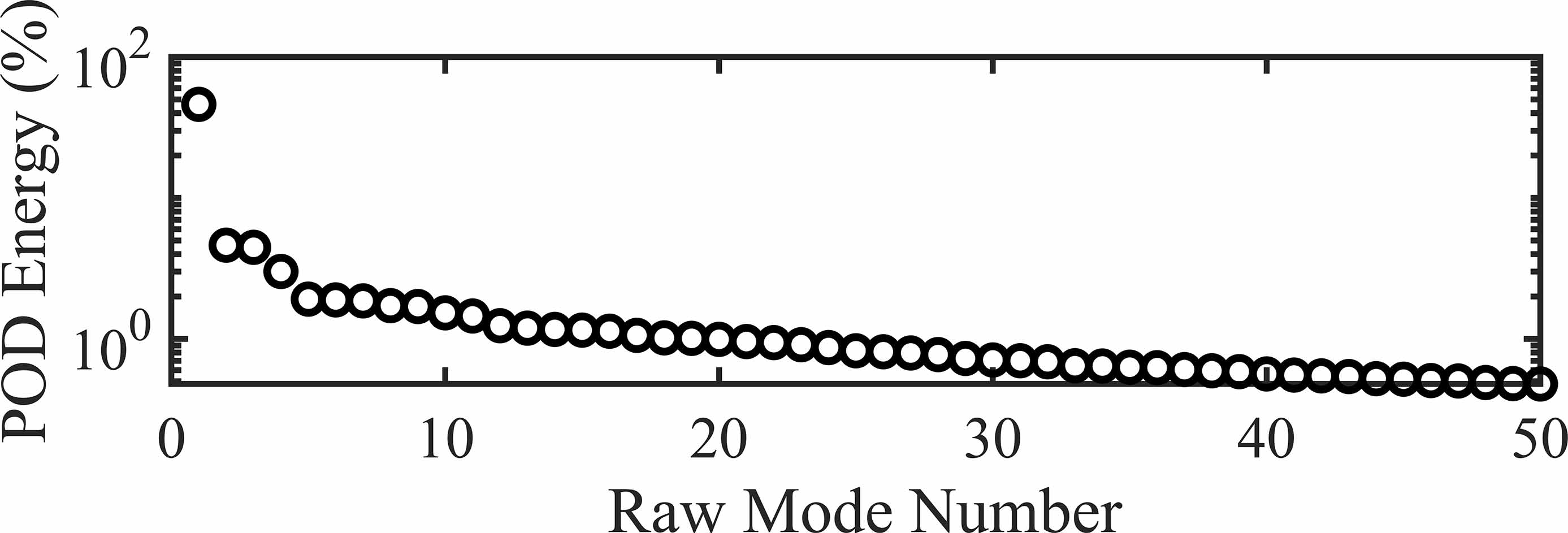}
         \caption{}
         \label{fig:break pod dist}
     \end{subfigure}
     
     \begin{subfigure}[b]{2.8in}
         \centering
         \includegraphics[width=\textwidth]{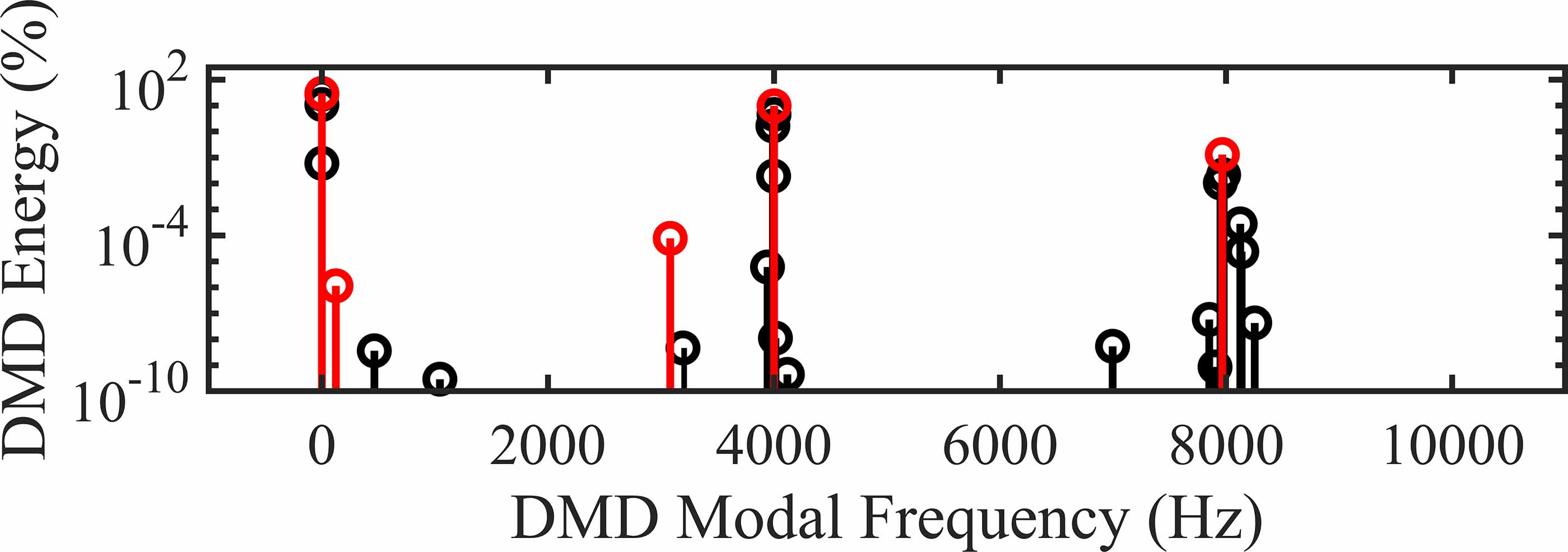}
         \caption{}
         \label{fig:break dmd spec}
     \end{subfigure}
     \end{minipage}
        \caption{(\subref{fig:break pod spec}) POD PSD, (\subref{fig:break pod dist}) POD energy distribution, and (\subref{fig:break dmd spec}) DMD spectrum for the jet breakup case.}
        \label{fig:break graphs}
\end{figure}

The modal structures are given in figure~\ref{fig:break modes} and accompanying modal information is given in figure~\ref{fig:break graphs} for the jet breakup case. Again, as system complexity is increased, the appropriate modal truncation becomes less clear through the POD energy distribution in figure~\ref{fig:break pod dist} as now the energy content of the first 50 modes are all within one order of magnitude. However, both the POD energy distribution and the DMD spectrum suggest three modes may dominate this system.

The second POD and DMD modes correctly capture the perturbation frequency and agree on the spatial structure. This mode can be segmented into three parts. The top part, just prior to droplet formation, correlates well with the pure dilational structures identified in the second mode in figure~\ref{fig:sin dmd modes} which has already been discussed. This indicates that this segment primarily undergoes a dilational process rather than jet breakup. The bottom segment identifies clear regions and shapes of formed droplets indicating that jet breakup has generally occurred further upstream. In between these two sections, however, there is a bifurcation of these droplet regions, as if representing a transition from the pure dilational behaviour to the pure droplet formation. This centre region has no pure physical meaning; instead the structures provide a \textit{probability distribution} of where liquid is present. From the video of this system, the location of jet breakup and consequent droplet formation fluctuates greatly within the captured field of view. This fluctuation is, in part, captured by the second mode as the centreline bifurcation of the formed droplets is caused by an uncertainty of whether the liquid column is still intact, but with a thinning of the jet, or has broken up within this region. Even the formed droplets in the bottom section are not true droplet shapes but instead indicate an uncertainty that liquid is present along the jet centreline.

Following on from the discussion of harmonics from the dilational jet case, based on the frequency content of the modes, the third DMD mode may be considered the second harmonic of the second mode. From before, this would suggest that the third mode only provides motion information of the fundamental spatial structures. However, the spatial structures between the modes are not just scaled by a factor of 2 but instead the third mode identifies new, unique spatial patterns. These finer spatial scales appear to capture the smaller satellite droplets which form above and below primary droplets, whose formation occurs on the jet centreline. In fact, these finer modal spatial scales almost exactly match the spatial scales of a satellite droplet for any time step. This mode still captures the motion of the primary droplets, whose spatial extent spans the width of the jet, but is interrupted by the presence of the satellite droplets. Higher harmonic modes, therefore, do \textit{not} necessarily give motion information only but can, in addition to motion information, provide new spatial information. The reason for the superposition of the motion and spatial information likely arises from the fact that the finer spatial scales captured by the third DMD mode aligns with the mean diameter of the satellite droplets. While a lot of information may be extracted from this third mode, it would be beneficial to separate the motion and spatial information content from this mode, as more complex systems may produce modes which are a superposition of numerous harmonics and spatial scales.

Similar to the previous findings for the POD modes, the third POD mode captures the low frequency side-to-side motion of the jet, while the next most dominant modes have no clear physical interpretation. The fourth and fifth DMD modes also do not have clear physical meaning. As a result, the jet breakup system should not be represented by these modes and, while other lower-energy and meaningful modes may still exist, a three-mode truncation is appropriate for this system.

\section{Analysis of jets in crossflow}\label{unknown}

\subsection{Column breakup in crossflow}

The laminar jet in crossflow is the simplest case to look at for the jet in crossflow application; a liquid column is broken up through hydrodynamic forces and aerodynamic forces induced by a gaseous crossflow. Extracted modes are expected to capture the flapping behaviour of the intact liquid column and the periodic breakup of the liquid jet, but all underlying processes and frequencies are now unknown.

\begin{figure}
    \centering
    \begin{subfigure}[b]{0.3\textwidth}
        \centering
        \includegraphics[height=4in]{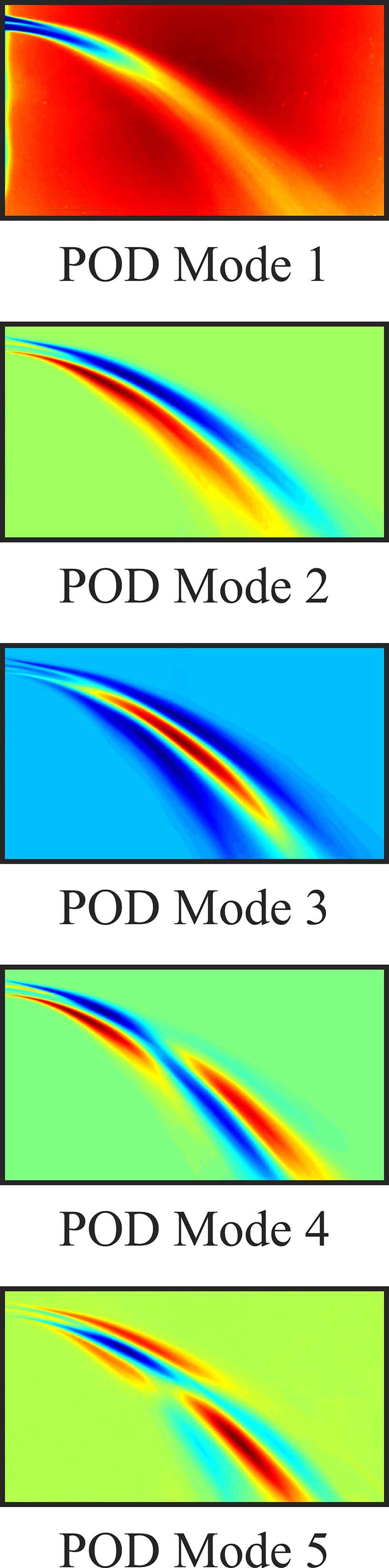}
        \caption{}
        \label{fig:lamjic pod modes}
    \end{subfigure}
    \begin{subfigure}[b]{0.3\textwidth}
        \centering
        \vspace*{0.1in}
        \includegraphics[height=4in]{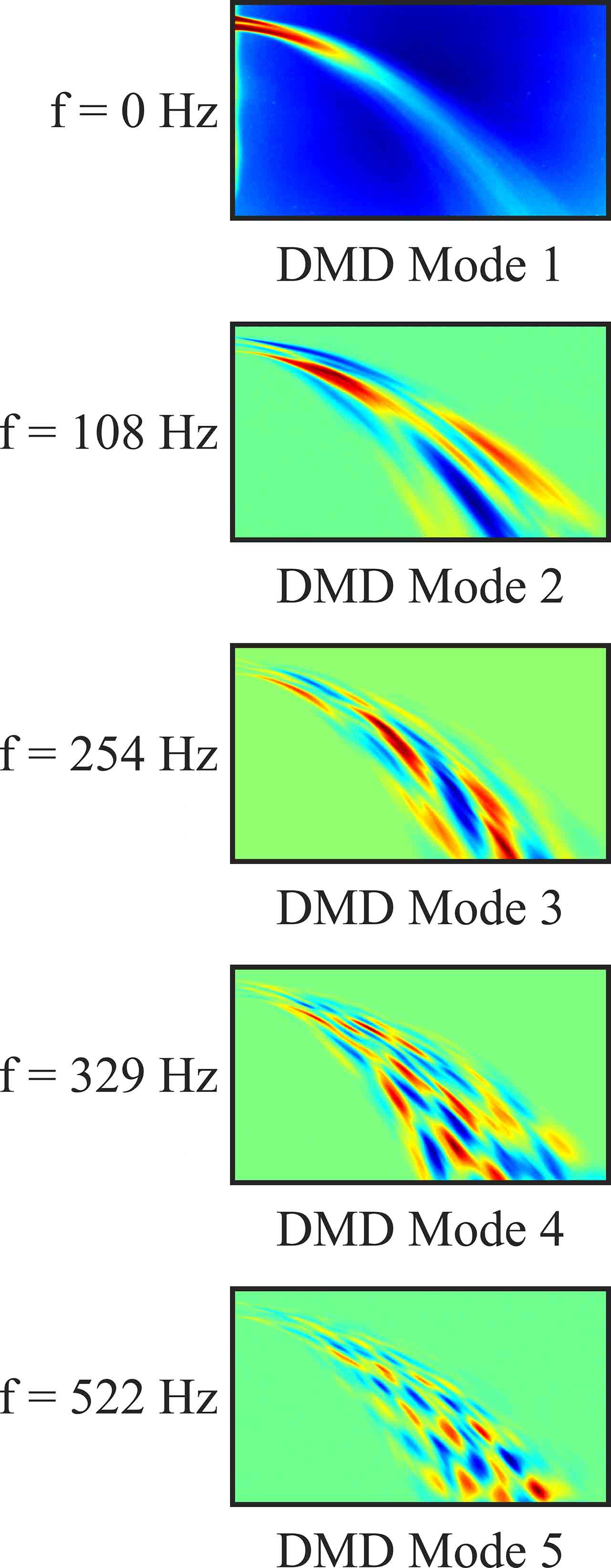}
        \caption{}
        \label{fig:lamjic dmd modes}
    \end{subfigure}
    \caption{The first five most dominant (\subref{fig:lamjic pod modes}) POD modes and (\subref{fig:lamjic dmd modes}) DMD modes for the column breakup regime.}
    \label{fig:lamjic modes}
\end{figure}

\begin{figure}
     \centering
     \begin{minipage}{.4\textwidth}
     \centering
     \begin{subfigure}[b]{1.5in}
         \centering
         \includegraphics[width=\textwidth]{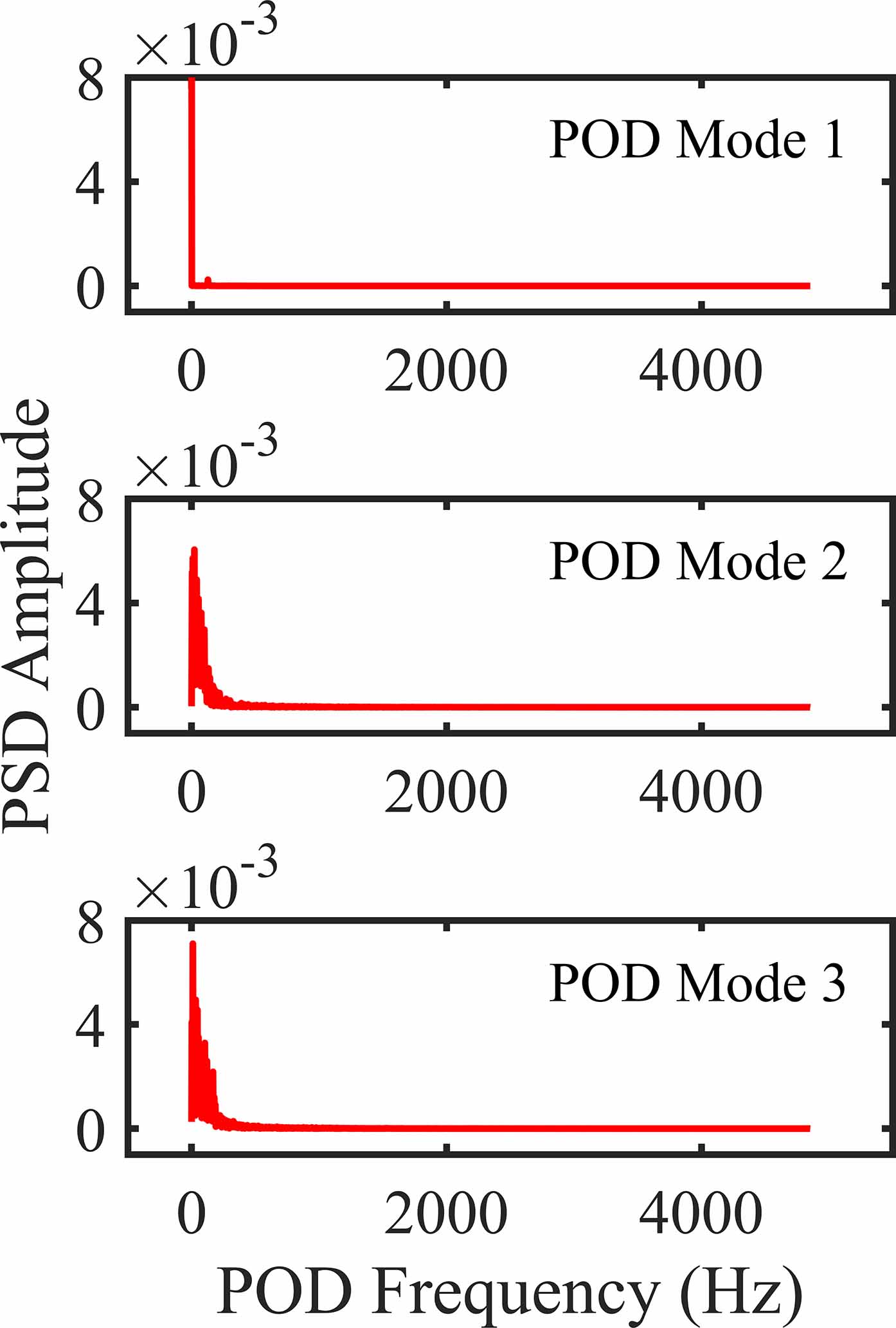}
         \caption{}
         \label{fig:lamjic pod spec}
     \end{subfigure}
     \end{minipage}%
     \begin{minipage}{.6\textwidth}
     \begin{subfigure}[b]{2.83in}
         \centering
         \includegraphics[width=\textwidth]{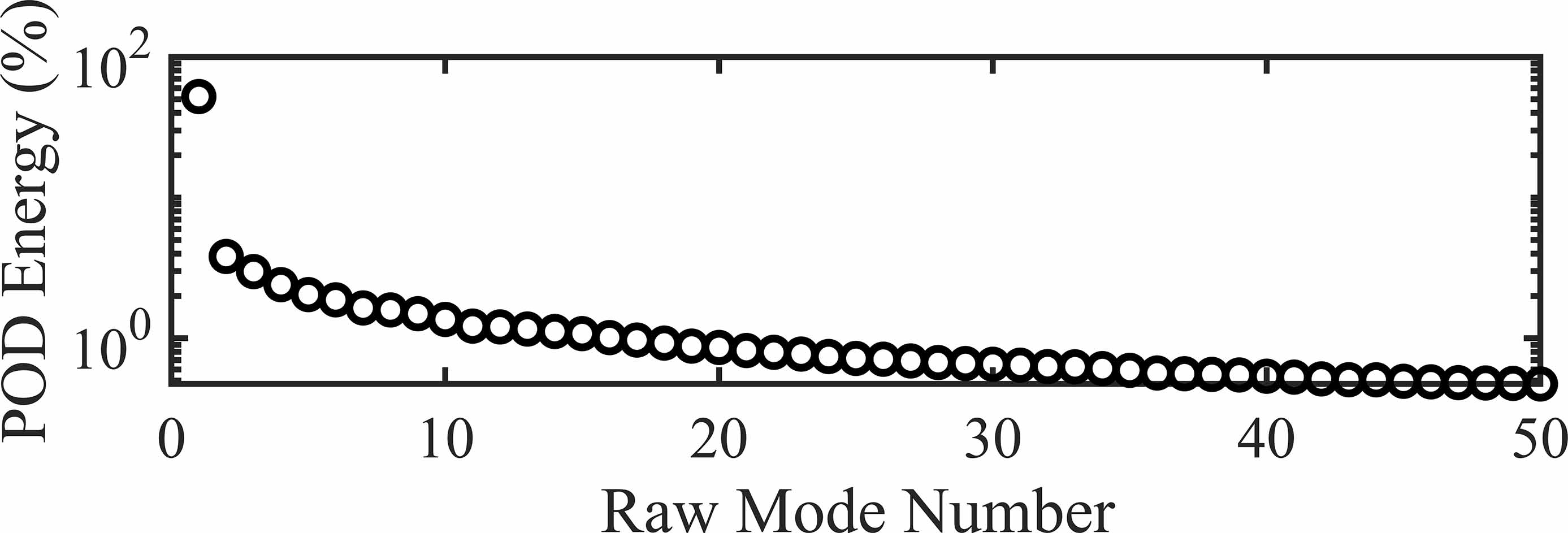}
         \caption{}
         \label{fig:lamjic pod dist}
     \end{subfigure}
     
     \begin{subfigure}[b]{2.8in}
         \centering
         \includegraphics[width=\textwidth]{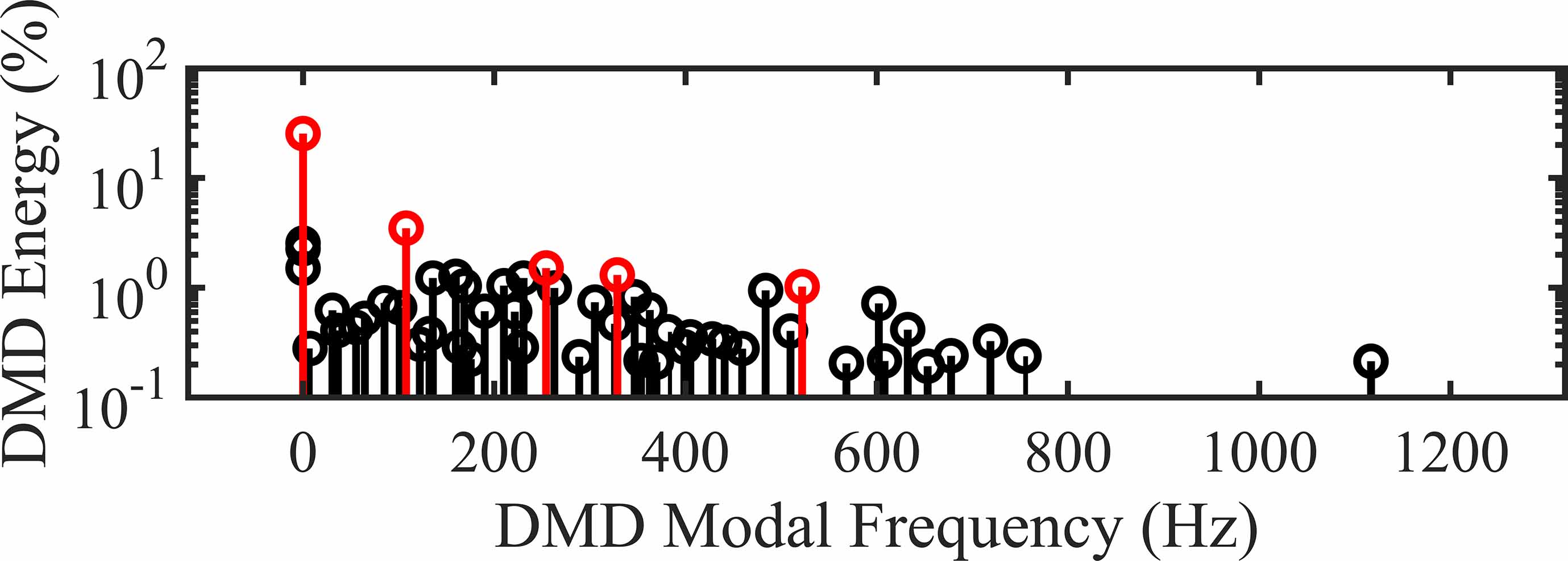}
         \caption{}
         \label{fig:lamjic dmd spec}
     \end{subfigure}
     \end{minipage}
        \caption{(\subref{fig:lamjic pod spec}) POD PSD, (\subref{fig:lamjic pod dist}) POD energy distribution, and (\subref{fig:lamjic dmd spec}) DMD spectrum for the column breakup regime.}
        \label{fig:lamjic graphs}
\end{figure}

The first five POD and DMD modes for the column breakup in crossflow case are shown in figure~\ref{fig:lamjic modes}. While for all of the simpler systems POD was able to capture some finer spatial features similar to DMD, for this simple jet in crossflow system POD is only able to capture modes with large spatial structures for the most dominant modes, and their corresponding PSD in figure~\ref{fig:lamjic pod spec} show a slight widening of the peak frequencies. Irrespective of the physical interpretation of the corresponding modes, a broader PSD does not as elegantly capture the dynamics of a system as a greater amount of information is required to represent the system. In contrast, DMD will maintain its spectral elegance for all systems analysed.

The second POD mode may capture the low frequency jet penetration depth fluctuation of the jet as the regions of maximal and minimal values agree well with the most regular jet trajectories. DMD does not produce a dominant mode with the same interpretation, similar to it being unable to capture the subtle side-to-side motion of the jets in the canonical flow cases. This may be an issue with the DMD energy formulation as in all of these cases DMD assigns less dominance to low frequency modes which capture large spatial scales.

The third POD mode may suggest a periodic thickening and narrowing of the jet, but this behaviour is not exhibited from the raw video of the system. The fourth POD mode has been found and discussed by Arienti and Soteriou \cite{arienti2009time} who interpret this mode as providing an extra bending degree of freedom to the column. This agrees with the flapping behaviour of the system, whereby the jet penetration depth fluctuates in time which induces a lag before the entire column changes trajectory, but performing a modal reconstruction does not provide any additional insight. This mode has a very similar structure as the fifth POD mode, where the latter includes an extra two spatial segments. The interpretation of these modes should likely be similar, but we have no clear way to validate what they represent.

A difficulty in interpreting the DMD modes is also found. Even selecting the most dominant modes is hindered for this relatively simple system as a broad range of frequencies are found to be present and dominant. The dominant modes are again selected based on a combination of their DMD energy and modal frequency. Here, we select modes at frequencies whose energy content are atypically high for its frequency. A general trend of decreasing modal energy with increasing modal frequency is shown, so dominant modes are selected when there is an increase in energy following a range of continuous decrease. This is most clearly shown for the fifth DMD mode which oscillates at 522~Hz, as modes from 350~Hz to 500~Hz are continuously decreasing until a new peak is found at 522~Hz.

We can improve our interpretation of the modes by considering the general nature of the laminar jet in crossflow. Although the jet penetration depth fluctuates in time, this fluctuation is primarily dependent on upstream changes in trajectory, after which the column follows a set path. This change in trajectory is captured by, for example, the fourth and fifth POD modes and the second DMD mode where a discontinuity exists in the maximal and minimal values along the flow path. At the point of discontinuity, the flow may continue along that trajectory (moving from a region of maximal values to a region of minimal values) and then the motion of the jet is captured, or the flow may change trajectory. The fifth POD mode may then identify initial jet trajectories, prior to the structural discontinuity, and then possible changes in trajectory due to the fluctuating penetration depth.

This reasoning can be extended towards the DMD modes, where the spatial extent of the structures increases with downstream distance due to the jet penetration depth fluctuation. As the DMD modal frequency increases, the spatial scales only reduce primarily in the direction of the flow which indicates the motion of the jet along a particular trajectory. Even for this simple jet in crossflow example it is difficult to validate the interpretation of these modes and to determine the legitimacy of the dominant modes as being meaningful. Using the findings derived from the canonical flows in tandem with analysing the raw data therefore becomes a necessary tool in interpreting these results.

\subsection{Bag breakup in crossflow}

\begin{figure}
    \centering
    \begin{subfigure}[b]{0.3\textwidth}
        \centering
        \includegraphics[height=4in]{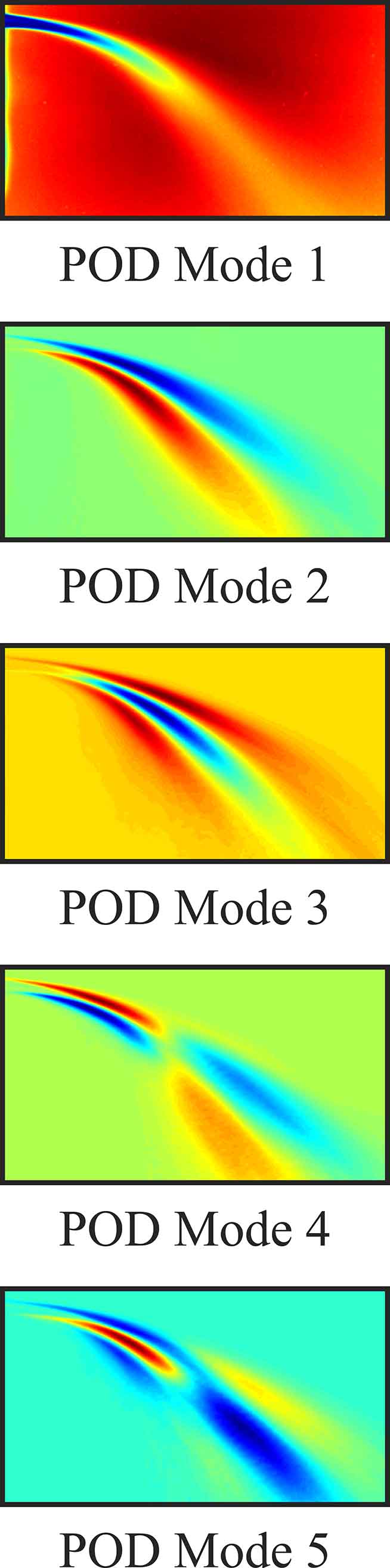}
        \caption{}
        \label{fig:bagjic pod modes}
    \end{subfigure}
    \begin{subfigure}[b]{0.3\textwidth}
        \centering
        \vspace*{0.1in}
        \includegraphics[height=4in]{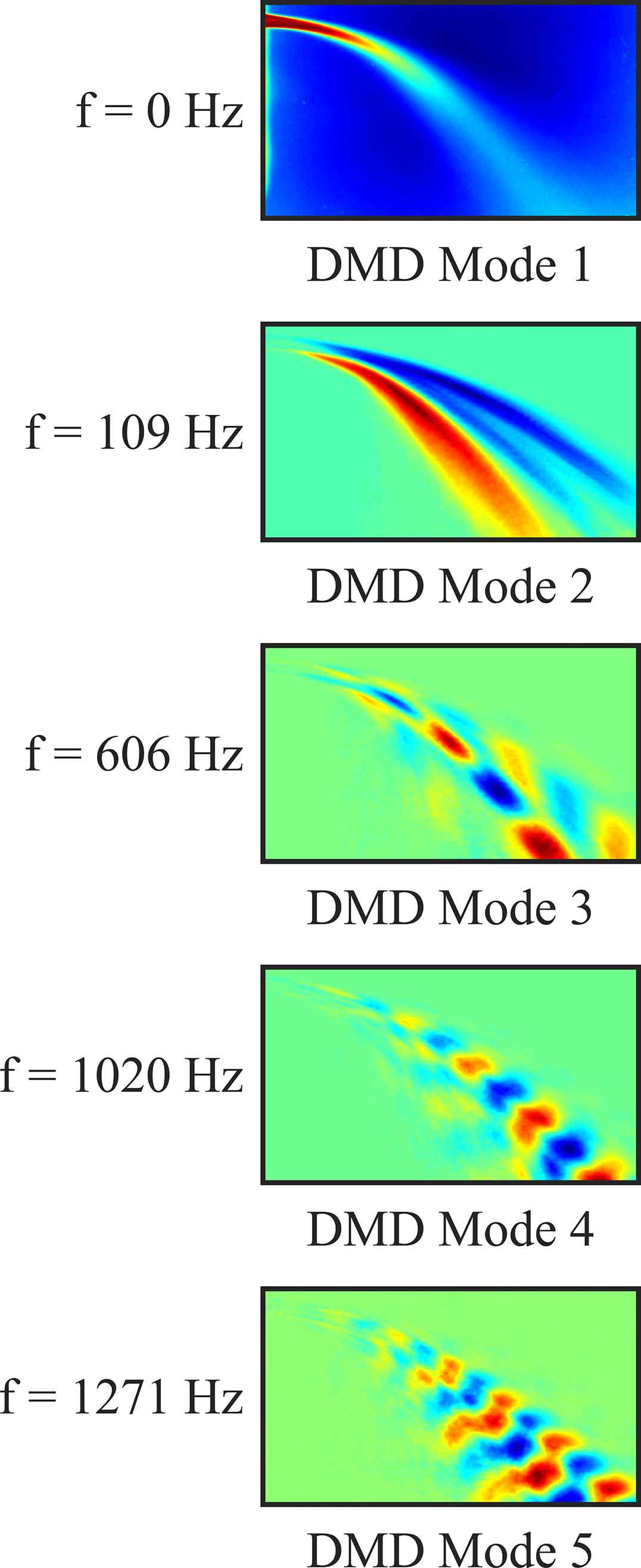}
        \caption{}
        \label{fig:bagjic dmd modes}
    \end{subfigure}
    \caption{The first five most dominant (\subref{fig:bagjic pod modes}) POD modes and (\subref{fig:bagjic dmd modes}) DMD modes for the bag breakup regime.}
    \label{fig:bagjic modes}
\end{figure}

\begin{figure}
     \centering
     \begin{minipage}{.4\textwidth}
     \centering
     \begin{subfigure}[b]{1.5in}
         \centering
         \includegraphics[width=\textwidth]{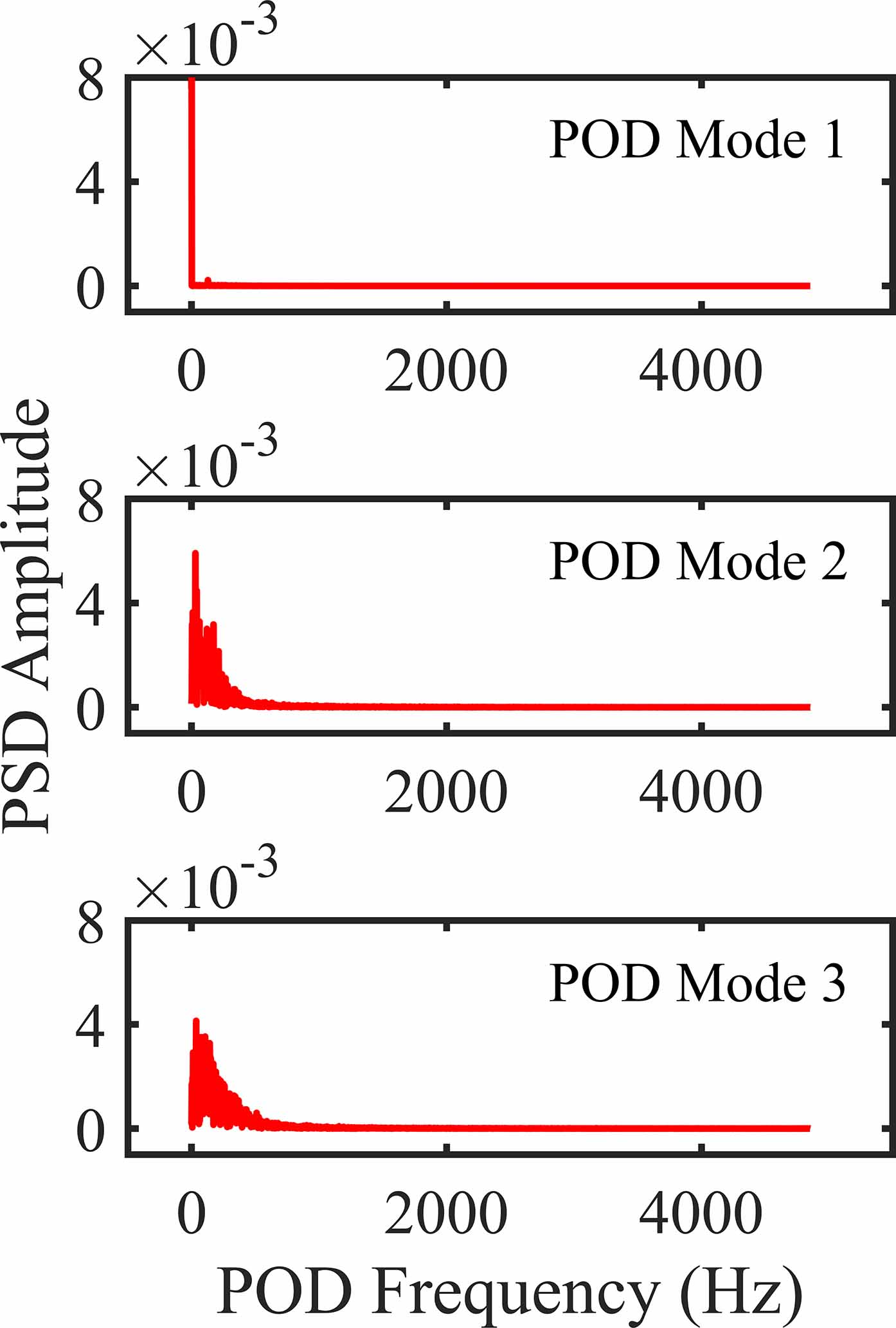}
         \caption{}
         \label{fig:bagjic pod spec}
     \end{subfigure}
     \end{minipage}%
     \begin{minipage}{.6\textwidth}
     \begin{subfigure}[b]{2.83in}
         \centering
         \includegraphics[width=\textwidth]{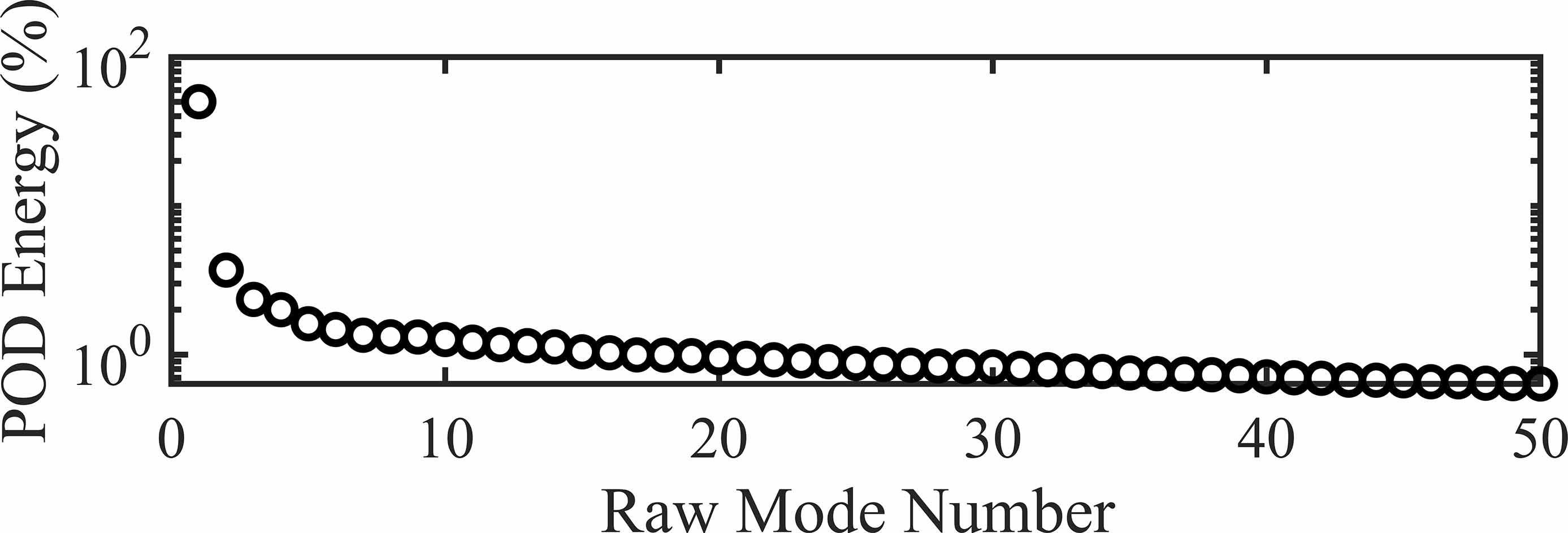}
         \caption{}
         \label{fig:bagjic pod dist}
     \end{subfigure}
     
     \begin{subfigure}[b]{2.8in}
         \centering
         \includegraphics[width=\textwidth]{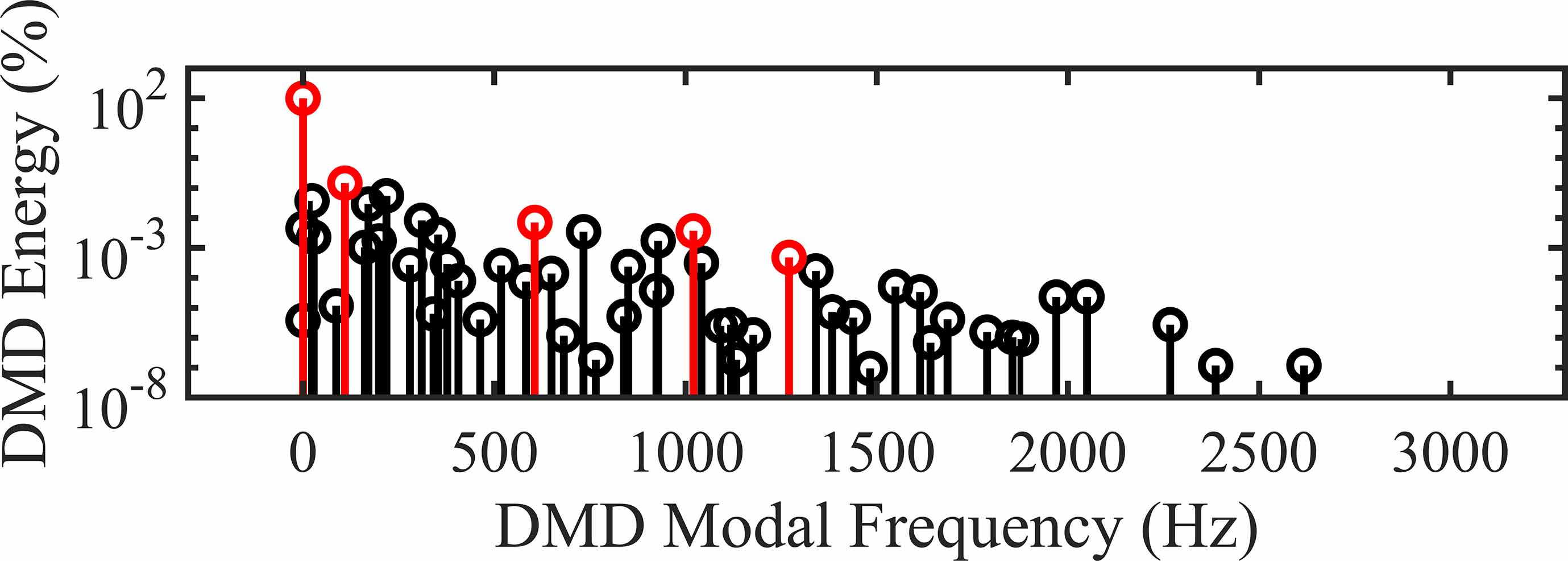}
         \caption{}
         \label{fig:bagjic dmd spec}
     \end{subfigure}
     \end{minipage}
        \caption{(\subref{fig:bagjic pod spec}) POD PSD, (\subref{fig:bagjic pod dist}) POD energy distribution, and (\subref{fig:bagjic dmd spec}) DMD spectrum for the bag breakup regime.}
        \label{fig:bagjic graphs}
\end{figure}

With an increased momentum flux ratio and aerodynamic Weber number, the column breakup from a crossflow transitions to the bag breakup regime, whereby segments of liquid are flattened and consequently expanded by crossflowing air to create a ``bag'' which then atomises. The extracted decomposition modes are provided in figure~\ref{fig:bagjic modes} and corresponding modal information are provided in figure~\ref{fig:bagjic graphs} for the bag breakup system.

Interestingly, the first five POD modes agree almost exactly with the POD modes for the column breakup case. The main difference is in a widening of the modal structures at further downstream locations as well as a further widening of the modal PSD peaks. While the latter may suggest a more complex system is being analysed, the modal structures indicate no \textit{high energy} difference between the bag and column breakup cases. This is not to say that all of the POD modes are identical between the two cases nor that lower energy modes contain meaningless information, rather it highlights the difficulty in systematically identifying modes which represent true underlying system processes. As POD is founded in energy-optimality, it does not make sense to manually identify lower energy modes to extract meaning regardless of what these findings suggest. Instead, a variant of POD would be preferred which produces high energy modes which capture distinguishing features between the two systems.

The second DMD mode captures the same low-frequency flapping of the jet as POD, where the flapping is again observed from the raw video data. This is further verified by identifying the periodic behaviour of the jet stream when it is initially located at either the maximal or minimal value locations. For this mode, this is done by comparing the general location of the jet every $f_s/f=9708/109\approx 89$ time steps which should match for most multiples of this value. This is indeed the case for this mode.

The third DMD mode has more concentrated spatial features towards the centreline of the mean jet flow. For these more complex systems especially, we borrow the findings from \S\ref{jet breakup} where we treat these modal structures as \textit{probability distributions} rather than absolute locations where liquid is present. This can complicate the interpretation of these modes so it is vital that reference is made back to the raw data to verify any interpretation made. Less prominent, phase-offset structures are at either side of the prominent centreline structure of the third mode. The flow which is preferentially found at the jet centreline is found to either buckle or undergo bag breakup at the spatial scales given by the third mode. The buckling tends to create segments which penetrate further into the crossflow which is associated with the less prominent structures penetrating deeper. Similarly, the shallower, less prominent structures are associated with segments which are undergoing bag breakup. The phase-offset of these structures arise naturally from segments of the jet whose penetration depth either increases or decreases with respect to the jet centreline. Liquid exists between these segments which remain undisturbed along the initial trajectory.

The fourth and fifth DMD modes capture wave-like features which are commonly presented and discussed in literature. Based on previous findings, these modes do capture the motion of liquid segments but it may be erroneous to infer that these structures capture true underlying spatial scales as we have shown that finer scales may be given by modes which are higher harmonics of fundamental modes but carry no additional spatial structure information. Whether a mode is a harmonic of a fundamental mode or whether it captures spatial structure information may require scrutinisation of the raw data or a modal correlation study. The fifth DMD mode may be the second harmonic of the third mode based on the frequency and spatial scale difference. As this mode has a different structural topology compared to the third DMD mode, it may still capture relevant liquid spatial scales in addition to the motion information of the centreline jet trajectory of the third DMD mode. This can be verified with the raw video data where the buckling of the jet produces segments with scales on the same order as the structures in the fifth DMD mode.

\subsection{Multimode breakup in crossflow}

The final system to be analysed is the multimode breakup system which occurs when the momentum flux ratio and aerodynamic Weber number are increased further beyond the bag breakup mode. Liquid is atomised both through bag breakup, so we would expect to see similar modal results as from the previous test condition, and through shear breakup where small droplets are sheared off the injected jet. All previous understanding should assist in interpreting and understanding the modal results.

\begin{figure}
    \centering
    \begin{subfigure}[b]{0.22\textwidth}
        \centering
        \includegraphics[height=4in]{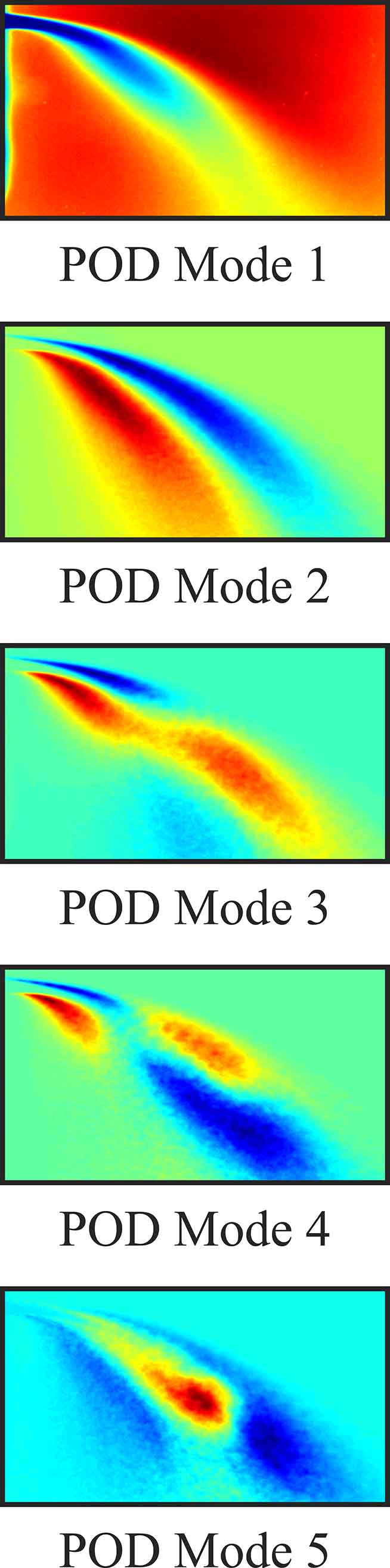}
        \caption{}
        \label{fig:multijic pod modes}
    \end{subfigure}
    \begin{subfigure}[b]{0.375\textwidth}
        \centering
        \vspace*{0.1in}
        \includegraphics[height=4in]{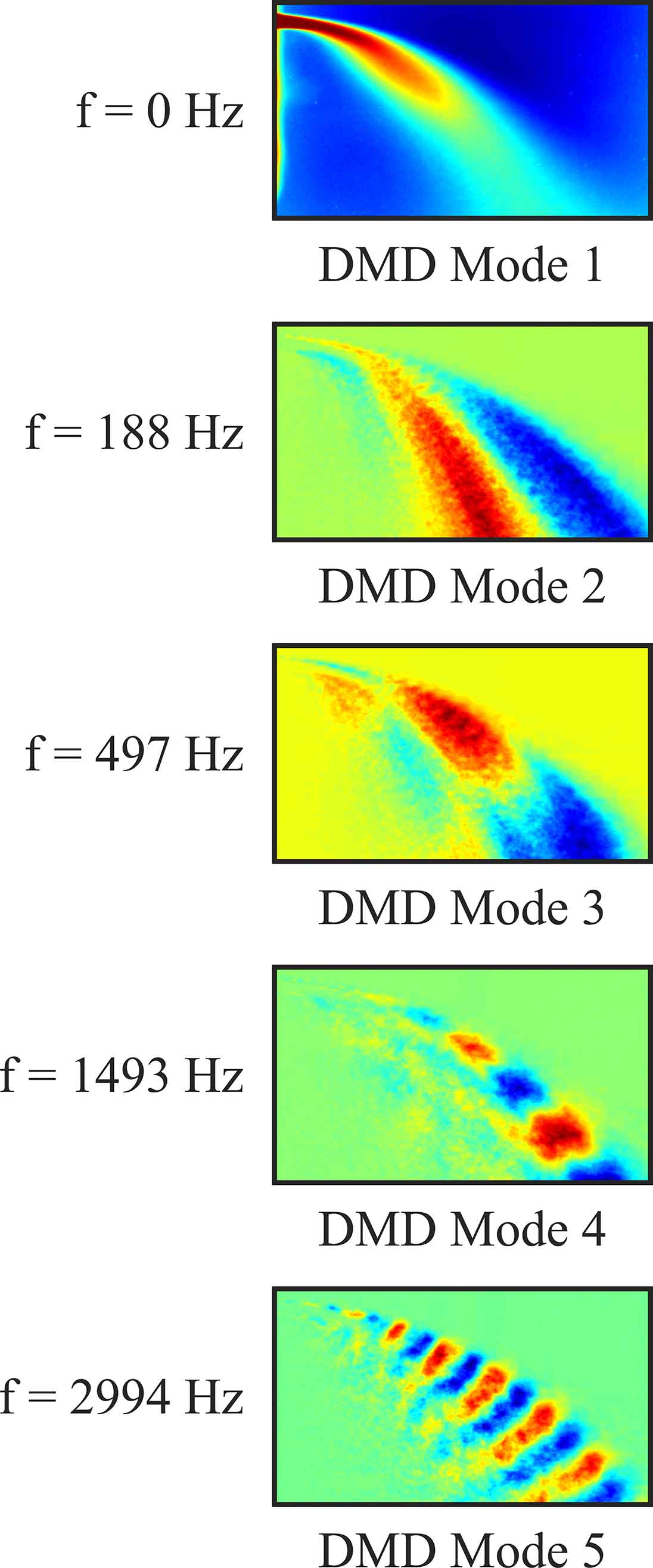}
        \caption{}
        \label{fig:multijic dmd modes}
    \end{subfigure}
    \begin{subfigure}[b]{0.375\textwidth}
        \centering
        \vspace*{0.1in}
        \includegraphics[height=4in]{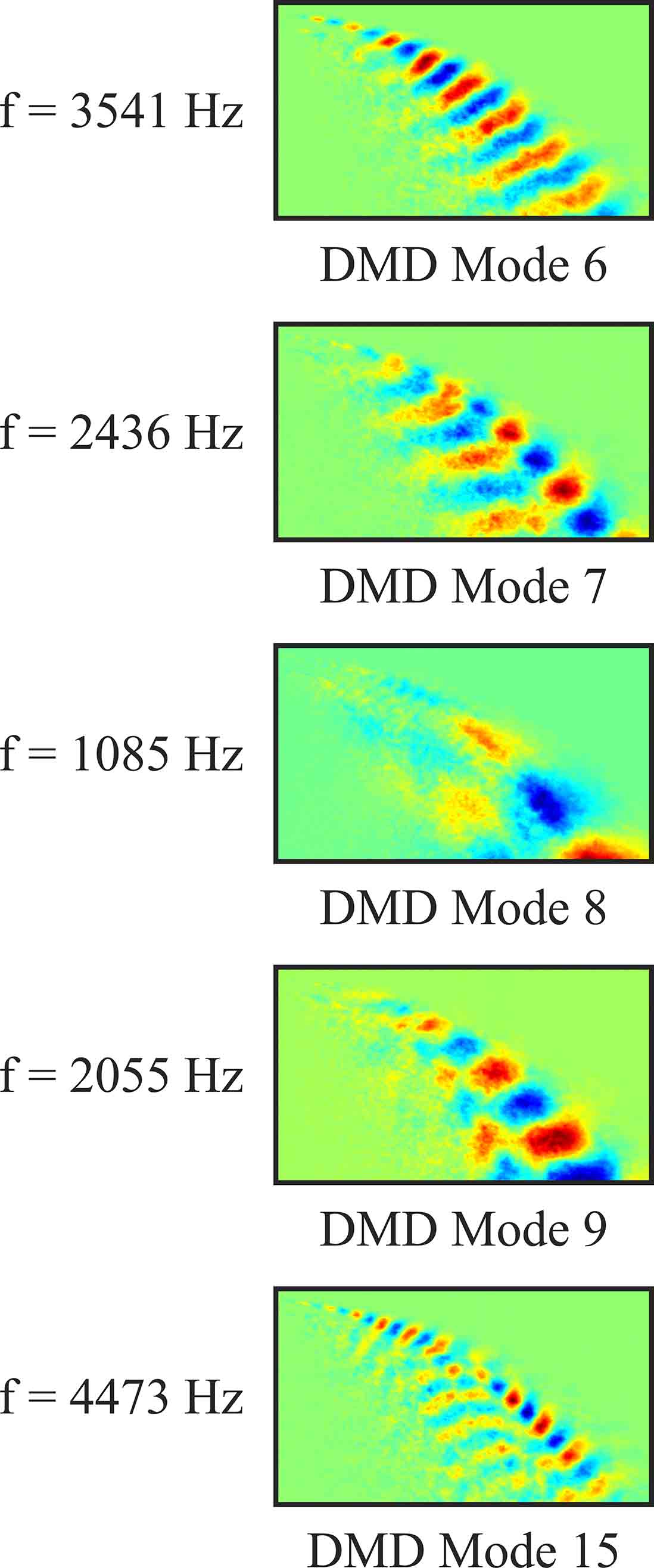}
        \caption{}
        \label{fig:multijic2 dmd modes}
    \end{subfigure}
    \caption{The first five most dominant (\subref{fig:multijic pod modes}) POD modes and (\subref{fig:multijic dmd modes}) DMD modes for the multimode breakup regime. (\subref{fig:multijic2 dmd modes}) The next five modes which may be harmonics of the third DMD mode.}
    \label{fig:multijic modes}
\end{figure}

\begin{figure}
     \centering
     \begin{minipage}{.4\textwidth}
     \centering
     \begin{subfigure}[b]{1.5in}
         \centering
         \includegraphics[width=\textwidth]{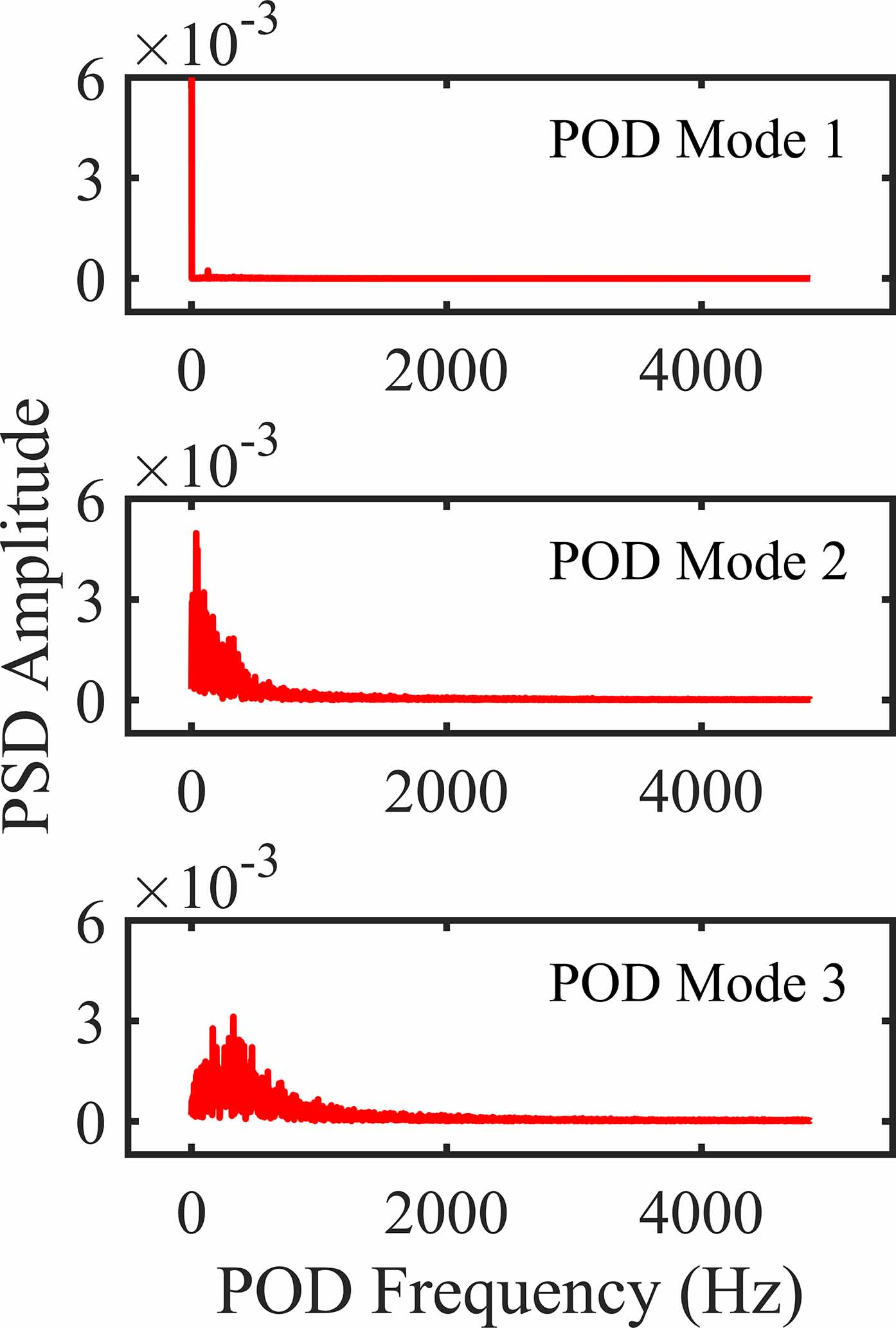}
         \caption{}
         \label{fig:multijic pod spec}
     \end{subfigure}
     \end{minipage}%
     \begin{minipage}{.6\textwidth}
     \begin{subfigure}[b]{2.83in}
         \centering
         \includegraphics[width=\textwidth]{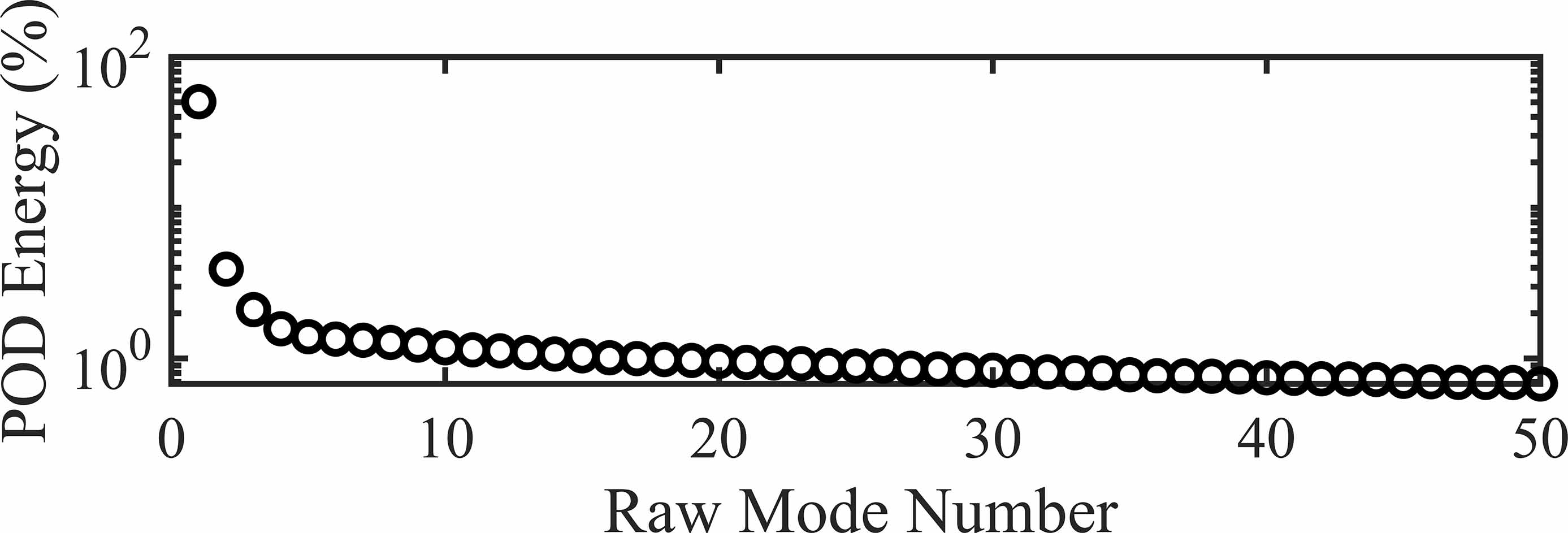}
         \caption{}
         \label{fig:multijic pod dist}
     \end{subfigure}
     
     \begin{subfigure}[b]{2.8in}
         \centering
         \includegraphics[width=\textwidth]{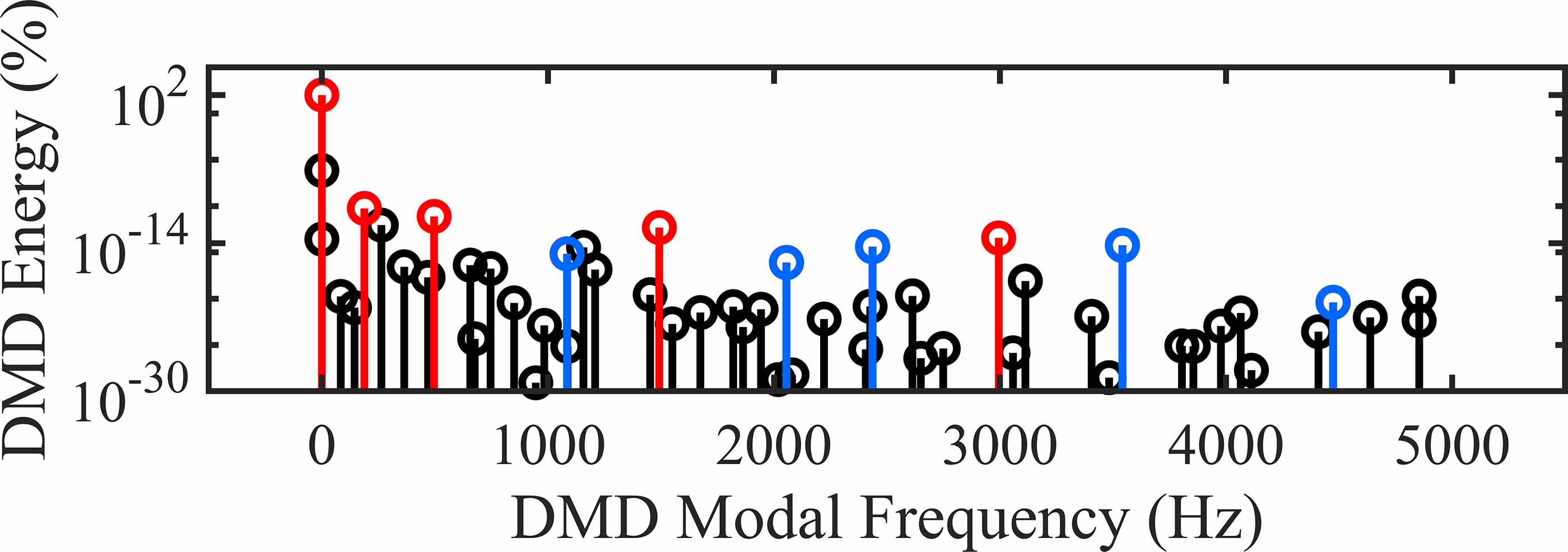}
         \caption{}
         \label{fig:multijic dmd spec}
     \end{subfigure}
     \end{minipage}
        \caption{(\subref{fig:multijic pod spec}) POD PSD, (\subref{fig:multijic pod dist}) POD energy distribution, and (\subref{fig:multijic dmd spec}) DMD spectrum for the multimode breakup regime.}
        \label{fig:multijic graphs}
\end{figure}

The modes and accompanying modal information for the multimode case are given in figure~\ref{fig:multijic modes} and figure~\ref{fig:multijic graphs}, respectively. The POD modes are again only capturing large scale structures which may capture the penetration depth fluctuations. The structures have increased size owing to the increased spatial distribution of the liquid from the shear breakup mode. Further, there is again a broadening in the PSD for the non-zero frequency modes indicating a more complex system. Although this may suggest a more complex system is being analysed, at least compared to the column and bag breakup modes, the most dominant modes for these systems are all very similar and offer little to provide salient features that distinguish the systems from each other.

The first five most dominant DMD modes are shown in figure~\ref{fig:multijic dmd modes}, which share some resemblance to the bag breakup DMD modes. Outside of the first two DMD modes, all of the modes presented here, including the less dominant modes given in figure~\ref{fig:multijic2 dmd modes} whose corresponding spectrum values are given by the blue stems in figure~\ref{fig:multijic dmd spec}, have spatial structures and frequencies related to the third DMD mode. It is possible that these captured modes are just higher harmonics of the third mode. If this is the case, then all of the modal structures which concentrate along the same trajectory may only give motion information about the jet. Therefore, the fourth, fifth, and sixth DMD modes do not give additional insight into the spatial structures or waves inherent in this system. For the other modes, particularly the seventh, ninth, and fifteenth DMD modes, a second, prominent region with a reduced jet penetration depth is present which may also capture spatial information at the corresponding spatial scales. In the case of the seventh and ninth DMD modes, where the two spatial trajectories present in these modes are phase-offset, they appear to capture bag breakup as periodic segments are expanded into bags with a reduced penetration depth. While this can be verified using the raw video data, there are a large range of spatial scales associated with bag breakup but these modes may capture the most prominent scales of interest.

For the finer scales of the fifteenth DMD mode, this mode may capture the shearing of smaller droplets from the jet while the jet's trajectory is mainly undisturbed. As the droplets produced by bag breakup are generally larger than the droplets formed by shear breakup, it makes sense that modal structures which are interpreted to being associated to these breakup regimes occur at different spatial scales. We note that the fifteenth mode was included as it appeared to be a direct harmonic of the third mode; we make no attempt to discuss and interpret all lower energy modes. However, it is possible that the decrease in DMD energy with increasing modal frequency occurs due to the shorter temporal scales of interest which capture less overall energy. A possible adaptation of the DMD energy metric is to normalise it by the temporal scale associated with the modal frequency. Doing this removes the penalty of modes with finer spatial and temporal scales that still capture meaningful information.

An attractive feature of DMD is in its ability to extract frequency content from a system and relate it to spatial structures. All DMD modes presented for this case provide a direct correlation between the size of the spatial structures captured and the corresponding frequency. As we have shown, these modes can capture motion information through the use of the modal frequency rather than pure oscillatory behaviour. Further, finer spatial scales in the system are only captured from high frequency modes as these finer features have relatively high energy and coherence over the temporal scale considered.

\section{Conclusions}

In this work, we have provided some example approaches in how to interpret the modes extracted from POD and DMD when applied to video data. Although this work was focused on liquid injection systems, we expect findings and intuitions gained here to also be beneficial in the analysis of other systems of interest as well as data in other formats. Further, while we have made an attempt to validate the interpretations presented, this is by no means exhaustive as we purposefully focused on only the most dominant modes as, in theory, these should represent the entire system optimally.

We provided analysis and interpretation of results for apparently trivial, canonical cases, where the underlying processes governing the fluid system are simple and largely known. Even for these cases, questions arise about how to systematically select the dominant modes and how to distinguish between true underlying processes and spurious processes. From these systems, however, we make insights into fundamental and harmonic modes where the former provides true spatial structures representative of the system and the latter provides motion information corresponding to the fundamental structures and \textit{may} provide additional spatial structures at finer scales. The ability to interpret these higher harmonic modes may be subject to a superposition of processes existing at varying spatial scales.

The understanding gained through the simpler systems was shown to improve the understanding and interpretation of more practical systems, however, validating these interpretations is difficult and requires understanding gained from the analysis of canonical flows in tandem with analysing the raw data. POD was found to be ineffective in identifying unique high energy modes across the jet in crossflow cases. Finding relevant low energy modes, while meaningful, bypasses the energy-optimal purpose of POD. In addition, as the complexity of the systems increase, there is an accompanying broadening of the modal PSD frequency bands which only complicates the modal interpretation. In contrast, we have shown that DMD maintains interpretability for even low energy modes by considering them as higher harmonics of more dominant fundamental modes.

While DMD is often used for its ability to extract system frequencies, we have demonstrated that modal frequencies can capture motion information rather than pure oscillatory behaviour. For the cases analysed in this work, the modal frequency should be considered an indicator of the temporal scales over which energy content and spatial structure coherence are relevant. This helps to explain why the modal spatial scales decrease with increasing modal frequency.

Finally, intuition into the theory has been provided to ensure that workers using these techniques are aware of their respective limitations and applications. While the theory is provided extensively in literature, discussion of the theory is often shrouded in abstraction to maintain generality which has the potential of inhibiting the interpretation of extracted results.

\bibliographystyle{unsrt}
\bibliography{references}

\end{document}